# Are wave union methods still suitable for 20 nm FPGA-based high-resolution (< 2 ps) time-to-digital converters?

**Wujun Xie [1], Haochang Chen [1] and David Day-Uei Li [1,] ***

[1] Faculty of Science, University of Strathclyde, Glasgow G4 0RE, United Kingdom; wujun.xie@strath.ac.uk (W.X.); haochang.chen@strath.ac.uk (H.C.); david.li@strath.ac.uk (D.L.)

* Correspondence: david.li@strath.ac.uk (D.L.)

Abstract: This paper presents several new structures to pursue high-resolution (< 2 ps) time-to-digital converters (TDCs) in Xilinx 20 nm UltraScale field-programmable gate arrays (FPGAs). The proposed TDCs combined the advantages of 1) our newly proposed sub-tapped delay line (sub-TDL) architecture effective in removing bubbles and zero-bins and 2) the wave union (WU) A method to improve the resolution and reduce the impact introduced from ultrawide bins. We also compared the proposed WU/sub-TDL TDC with the TDC combining the dual sampling (DS) structure and the sub-TDL technique. Moreover, we introduced a binning method to improve the linearity and derived a formula of the total measurement uncertainty for a single-stage TDL-TDC to obtain its root-mean-square (RMS) resolution. Results conclude that the proposed designs are cost-effective in logic resources and have the potential for multiple-channel implementations. Different from the conclusions from a previous study, we found that the wave union is still influential in UltraScale devices when combining with our sub-TDL structure. We also compared with other published TDCs to demonstrate where the proposed TDCs stand.

---

## 1. Introduction

A time-to-digital converter (TDC) can measure the time interval between two events with a high resolution. Combined with single-photon detectors [1,2], TDCs have been applied widely in clinical positron emission tomography (PET) [3,4], light detection and ranging (LiDAR), robotics and self-driving vehicles [5–8], fluorescence lifetime imaging microscopy [9–12], quantum communications [13,14], time-of-flight (ToF) imaging [15,16], and nuclear/particle physics [17,18]. TDCs can be essential in all-digital phase-locked loops, and digital synthesizers for other applications [19–22].

TDCs can be implemented in analog or digital methods. Boosted by recent advances in CMOS manufacturing technologies, digital TDCs with a sub-nanosecond resolution have become much more prevalent in highly integrated systems. Compared with the digital TDCs implemented with the application-specific integrated circuits (ASIC), field-programmable gate array (FPGA) based TDCs are: 1) fast and easy to prototype, 2) low cost, and 3) reprogrammable.

The resolution (also called the least significant bit, LSB, or the average bin size) of a TDC is the lowest time interval that it can measure. Many applications particularly require high-resolution (<10 ps) TDCs and the demands have been growing strongly significantly for applications in LiDAR [23,24], PET systems [3,4], and time-domain diffuse correlation spectroscopy [25–27].

The tapped delay line (TDL) architecture has been a mainstream approach [28–30] for FPGA-TDCs since carry-chain modules can easily construct TDLs in modern FPGA devices [31–35]. The propagation time of a signal propagating through a TDL can be measured and digitized. The resolution of a TDL-TDC is related to the signal propagation delay in TDLs.

Compare with previous FPGA structures, CARRY8 modules in Xilinx UltraScale FPGAs can double the number of taps of a TDL. With this structure, the resolution of a double sampling (DS) TDL-TDC achieved 2.25 ps [36]. To further improve the resolution of FPGA-TDCs, the Vernier delay line (VDL), the multi-phase design and the multi-chain design have been proposed to overcome process-related limitations [37–47]. These methods can achieve a better resolution than raw TDL architectures. However, these methods consume logic resources significantly, and the systems are complicated. The stochastic method, exploiting the stochastic properties of a set of latches to achieve a higher resolution, is popular in ASIC TDC designs but cannot be easily applied to FPGA designs. In [42], a 2D semi-stochastic Vernier structure was proposed to enhance the performance.

Some researchers used FPGA digital signal processing (DSP) blocks to build a TDL [48,49]. However, the linearity is still low. The linearity of a TDC is also a critical parameter to be ensured; a low-linearity TDC usually causes severe measurement uncertainties, even with a high resolution. The linearity performances of TDCs can be characterized by the differential nonlinearity (DNL) and the integral nonlinearity (INL) [50]. DNL and INL are defined as:

$$DNL[i] = \frac{W[i]-W_{ideal}}{W_{ideal}}, \tag{1}$$

$$INL[i] = \sum_{n=0}^{i} DNL[n], \tag{2}$$

where $W[i]$ is the width of the $i$-th bin and the $W_{ideal}$ is the ideal bin width. The DNL is the deviation of a quantization step from the ideal value of 1 LSB, and the INL is the accumulation of the DNL. We can obtain DNL and INL from code density tests.

About FPGA-TDCs, the non-uniformity of carry-chains and clock skews are two main reasons for nonlinearity [44]. The clock skew is caused by the dedicated clock distribution tree in FPGAs, and large clock skews usually appear at the boundaries or in the middle of clock regions (CRs). The non-uniformity of carry-chains deteriorates the linearity of entire delay lines and generates ultra-small bins and ultra-wide bins. The ultra-small bins (DNL $\leq$ -0.90 LSB) can be merged, and the zero-width bins (DNL = -1.00 LSB) need to be abandoned (since there are unable to capture any information). However, ultra-wide bins deteriorate the precision of measurements significantly. Wu *et al.* proposed the wave union (WU) method to ease ultra-wide bin problems and improve the precision [51]. A wave union signal, containing several rising edges (0-1 transitions) and falling edges (1-0 transitions), performs multiple measurements with a single TDL. The ultra-wide bins are sub-divided with this method.

To further improve the linearity and precision, calibrations are usually needed in TDCs. Previously published calibration techniques include the bin-by-bin calibration [31,52–54] and bin-width calibration techniques [55,56].

Furthermore, the mismatches between carry-chains in FPGAs lead to bubble problems [38] and cause encoding failures [35]. Traditional de-bubble operations use logic gates to remove bubbles or to recognize real signal transitions [38,54,57]; however, they introduce extra logic resources and clock cycles [38]. Bubble problems become more severe in more advanced UltraScale FPGAs. Moreover, the de-bubble method proposed in [58] performs differently for the 0-1 and 1-0 transitions. Therefore, the bubble problems persist, prompting Wang and Liu to conclude that the wave union method is not suitable for UltraScale FPGAs [36]. Since then, there is no efficient WU TDC implemented in UltraScale FPGAs reported. However, in 2018, the decomposition [35] and our sub-TDL [55] methods were proposed to more efficiently remove bubbles and zero-width bins without consuming extra logic resources.

Chen and Li reported that with sub-TDL structures, bubbles could be removed entirely [55]. It prompted us to ponder: A) whether the WU method still cannot be employed in UltraScale FPGA devices as suggested by [36] if bubble-resistant sub-TDL structures are available; B) whether the advantages of the sub-TDL design can be applied to DS structures, given the fact that the new CARRY8 modules contain both carry C and sum S output ports for the DS structure [36]; C) whether a combination of the above strategies can be exploited to achieve a resolution (towards 1 ps LSB conversion) with maintained linearity. Although UltraScale FPGAs were first introduced in 2014, there are few efficient TDCs reported in UltraScale FPGAs [32,36,55]. Migrating existing methods to such advanced FPGAs is not trivial at all, as the carry chains are slightly different from earlier generations. Moreover, a detailed uncertainty analysis has not been reported. This study aims to investigate the suitability of the WU method in UltraScale FPGAs and to alleviate related problems.

With the above rationales, the main innovations of this work include:

1) We clarified the difference in the propagation speed between the rising edge (0-1 transition) and the falling edge (1-0 transition) of the WU signal with an equation and a figure providing quantitative information. We reached a new conclusion different from a previously published study [36] about the suitability of the WU method in UltraScale devices; our analysis shows that the WU method is still powerful if it integrates our recently proposed sub-TDL structure [55]. We presented the first efficient single-TDL WU TDC in 20 nm UltraScale FPGAs. The outcomes should be able to convince researchers within the community to continue applying the WU method in more advanced FPGAs.

2) We implemented the first efficient WU TDC in UltraScale FPGA devices by integrating our newly proposed sub-TDL structure [55]. To present its efficiency, we also compared it with three other different TDC structures. Four TDC structures employing sub-TDL designs were proposed, implemented, and tested in the Kintex UltraScale KCU105 Evaluation Kit (UltraScale XCKU040 FPGAs):

a) WU/Sub-TDL TDC, denoted as WU TDC,
b) DS/Sub-TDL TDC, denoted as DS TDC,
c) DS/WU/Sub-TDL TDC, denoted as DSWU TDC,
d) Binned-DSWU TDC with a proposed binning method to improve the linearity.

The study further demonstrates that the WU method promises an efficient solution.

3) Detailed analysis of measurement uncertainties and error sources for the four proposed TDCs have been conducted.

## 2. Design and Architecture

For the proposed methods, the TDL architecture was used and constructed with cascaded carry chains CARRY8 and implemented in the UltraScale XCKU040 FPGA. Figure 1 shows the block diagrams of the proposed four TDCs. The DSWU and binned DSWU TDCs are derived from the combination of the structures shown in Figures 1(a) and 1(b). The DS TDC is the combination of the designs shown in Figures 1(b) and 1(c), whereas the WU TDC is constructed by the structures shown in Figures 1(a) and 1(d). We employed the newly proposed sub-TDL architectures in all four TDCs. Figure 1(a) shows a WU launcher generating a WU signal (containing a rising edge and a falling edge to provide additional sampling) into a TDL. The delay taps are sampled by the Sub-TDL Rising and Sub-TDL Falling modules. For non-WU designs shown in Figure 1(c), the hit signal is fed to the TDL directly. The launcher and related modules, such as the sub-TDL and the encoder for the falling edge, are removed. From Figures 1(a) and 1(c), the signal width from the CARRY8 depends on whether the DS method is employed. With the DS method, all outputs (both carry C outputs and sum S outputs, 16 bits in total) of the CARRY8 are sampled by a sub-TDL module, as shown in Figure 1(b). However, for the non-DS designs, only C outputs of the CARRY8 (8 bits in total) are considered (see Figure 1(d)). Note that there is still a buffer between S and C ports. In CARRY8, this buffer is a 2-to-1 multiplexer [59]. With the sub-TDL architecture [55], the raw thermometer code generated by a TDL is split into several subsets. These sub-thermometer codes are converted to corresponding one-hot codes (by TM2OH) and then encoded to binary codes (by OH2BIN) in the encoding module. The fine code of the TDC is the sum of these binary codes. From Figure 1, we conclude: 1) The numbers of the sampled taps from a CARRY8 module are $8 \times 2 \times 2 = 32$ (sampled by all outputs of the CARRY8 and then sampled sequentially by the Sub-TDL Rising and Sub-TDL Falling modules) for DSWU or binned DSWU TDCs or $8 \times 2 = 16$ for DS or WU TDCs. A TDC using neither WU nor DS methods only has eight taps sampled from a CARRY8 module, as shown in Figure 1(d). Architectures with more sampled taps offer a higher time resolution. 2) The sub-TDL structure can remove bubbles.

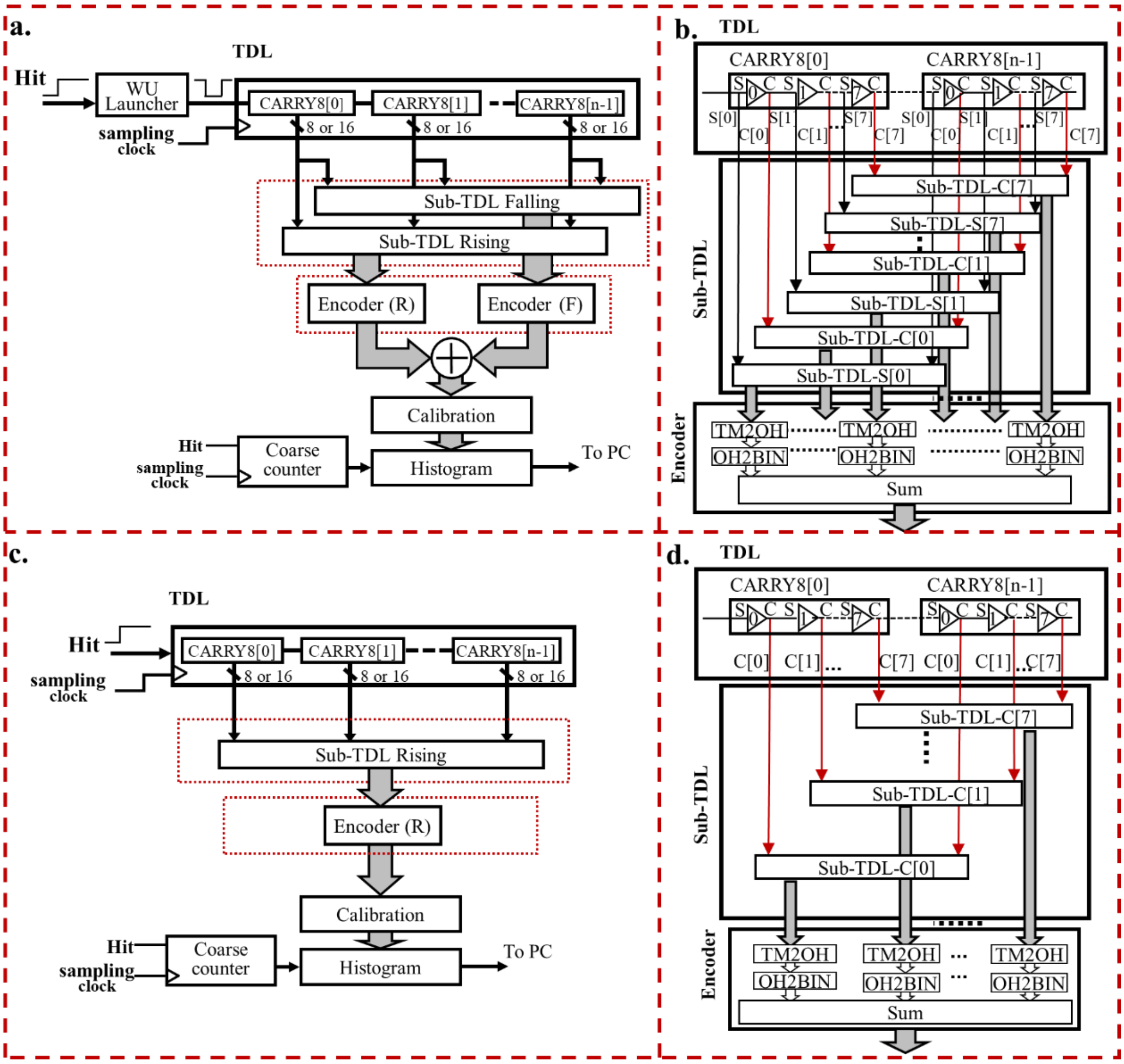

**Figure 1.** (a) Block diagram for the proposed TDC systems with the WU method. (b) Block diagram for the sub-TDL structure with the DS method and the encoder. (c) Block diagram for the proposed TDC system without the WU method. (d) Block diagram for the sub-TDL structure without the DS method and the encoder.

### 2.1. Dual-sampling Structure with Sub-TDL

The DS TDL structures were proposed in [32,36]. Compared with previous FPGA structures, CARRY8 structures in UltraScale FPGAs double the number of the taps or the equivalent bins of a TDL with the same total propagation delay. Therefore, the theoretical LSB or the average bin size can be reduced. However, the linearity degrades when the resolution is enhanced. A DS TDC with the sub-TDL encoding structure [55] has been proposed in this study. Figure 2 presents the linearity performance for the proposed DS TDC. This TDC achieved 2.53 ps resolution, and the length and the location of the TDL are confined within a central clock region.

In Figure 2, the broader bins around bin 450 cause a distinct step in the INL curve. Similar phenomena can be observed in the proposed DS and DSWU TDCs. The reason for these phenomena is that significant clock skews exist due to the bifurcation of the clock distribution tree in the middle of a clock region. To overcome this problem and to correct the accumulated offset, bin-by-bin remapping or mixed-calibration can be used. The main idea of the bin-by-bin calibration is to calibrate the times to the centers of the bins to correct the INL. Besides, the double-phase sampling architecture [44] can be used to halve the length of the TDL to avoid the clock skew problem. Figure 3 presents the linearity performance for the DS TDC with a range of 390 bins, and the $\mathrm{INL}_{pk-pk}$ (peak-to-peak INL) is improved to 9.80 LSB.

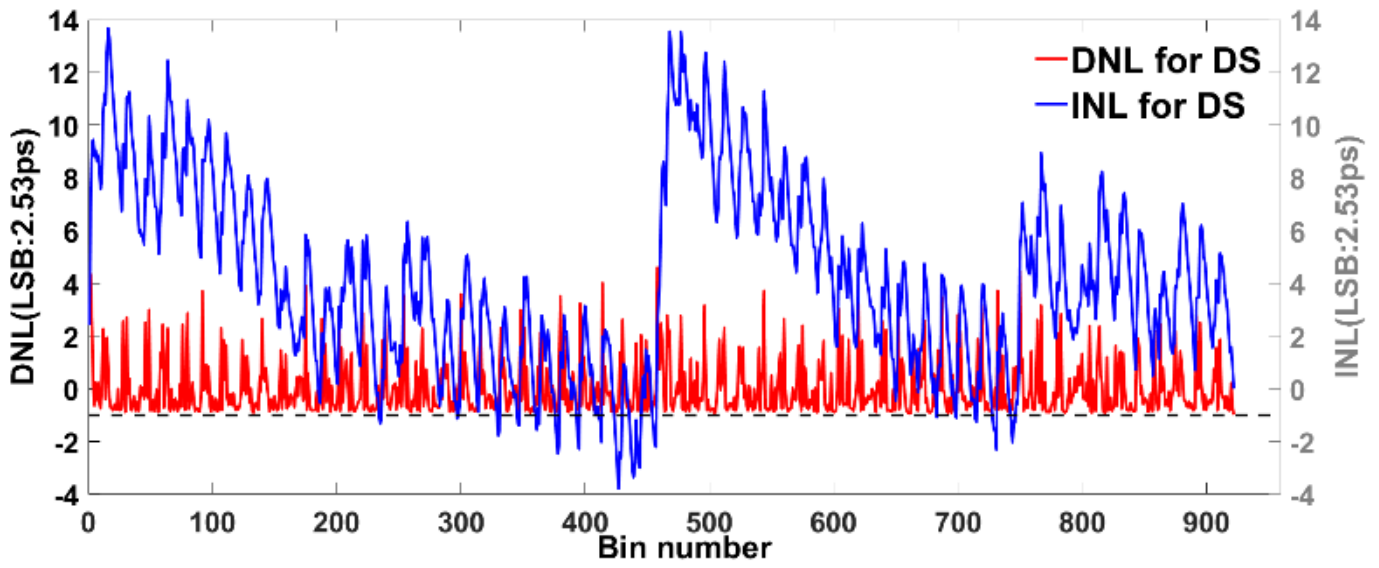

**Figure 2.** DNL and INL curves for the DS TDC.

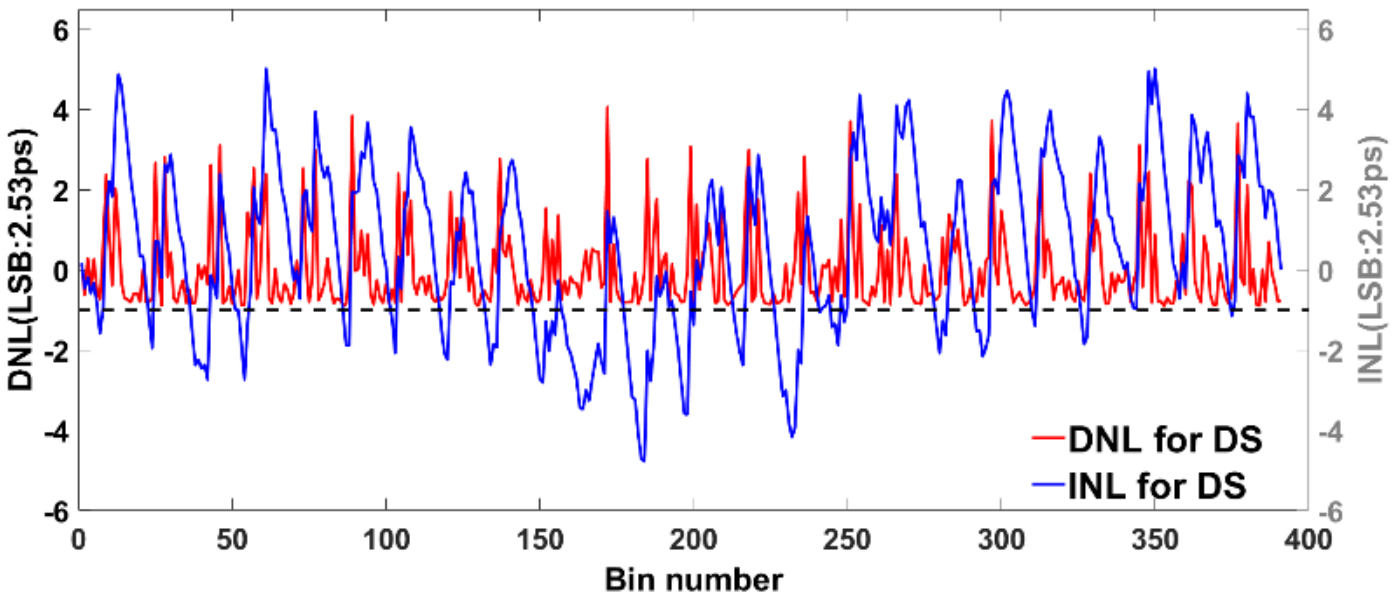

**Figure 3.** DNL and INL curves for the DS TDC (only using 390 bins).

### *2.2. Wave Union Method*

The WU method is another way to improve the resolution and alleviate ultra-wide bin problems. As shown in Figure 4, with the WU signal containing the rising and falling sampling edges, a WU TDC can perform like a TDC with two delay lines, one for the rising edge and another for the falling one. The LSB of the WU TDC is:

$$LSB_{wu} = \frac{MR}{N_{wu}} = \frac{MR}{N_{rising}+N_{falling}} = \frac{LSB_{rising} \times LSB_{falling}}{LSB_{rising}+LSB_{falling}}, \quad (3)$$

where $MR$ is the measurement range of a simple delay line. $N_{wu}$, $N_{rising}$, and $N_{falling}$ are the bin numbers of the WU TDC, the plain TDC with the rising edge and the plain TDC with the falling edge, respectively. The WU method can divide the ultra-wide bins in each raw measurement efficiently, and the average of these measurements yields a finer TDC resolution [60].

An effective launcher, capable of producing a stable WU signal, is vital for the whole system. The jitter caused by the WU signal, however, usually feeds to the TDL, and degrades the linearity.

The WU method has been studied by serval groups [52,53,58,61]. However, the propagation speeds of the falling edge and the rising edge are different in Kintex-7 FPGAs [58]. The WU interpolation efficiency on the resolution was revealed in a study, performed in different FPGAs, from 65 nm to 28 nm processes [62]. The results also indicated that the speed of the rising edge is different from that of the falling edge.

We have performed similar tests in UltraScale FPGAs. The results do show the difference in the propagation speed between the falling sampling edge and the rising sampling edge. Figure 5(a) shows the principle of the propagation speed difference between the rising edge and the falling edge; the gap between the rising edge and the falling edge varies when the WU signal propagates along the TDL. Figure 5(b) shows the relationships between the bin number and the measured time for the b.

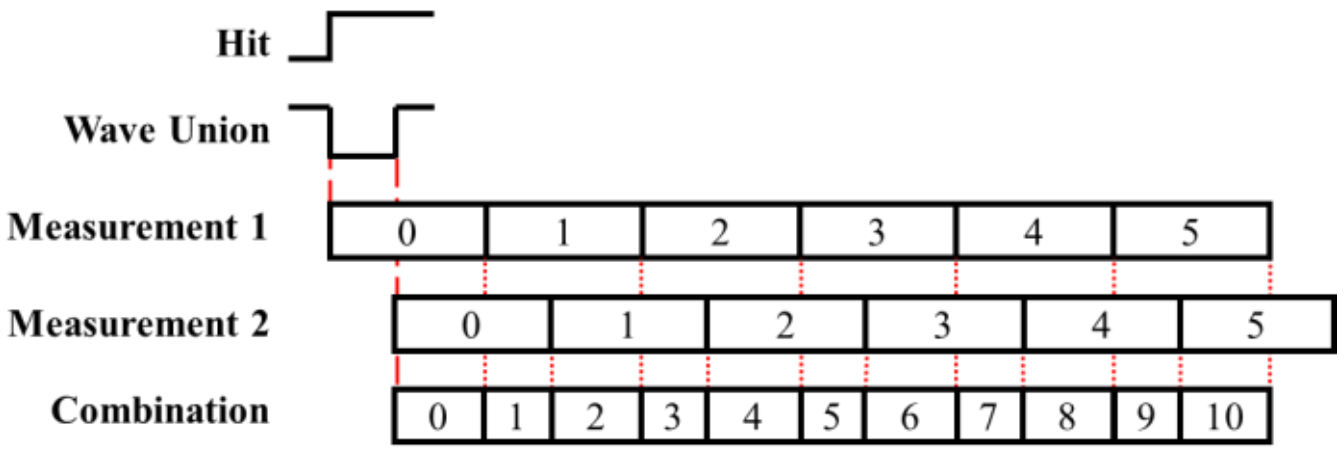

**Figure 4.** Concept of the WU method. The wave union only contains a pair of 0-to-1 and 1-to-0 transitions

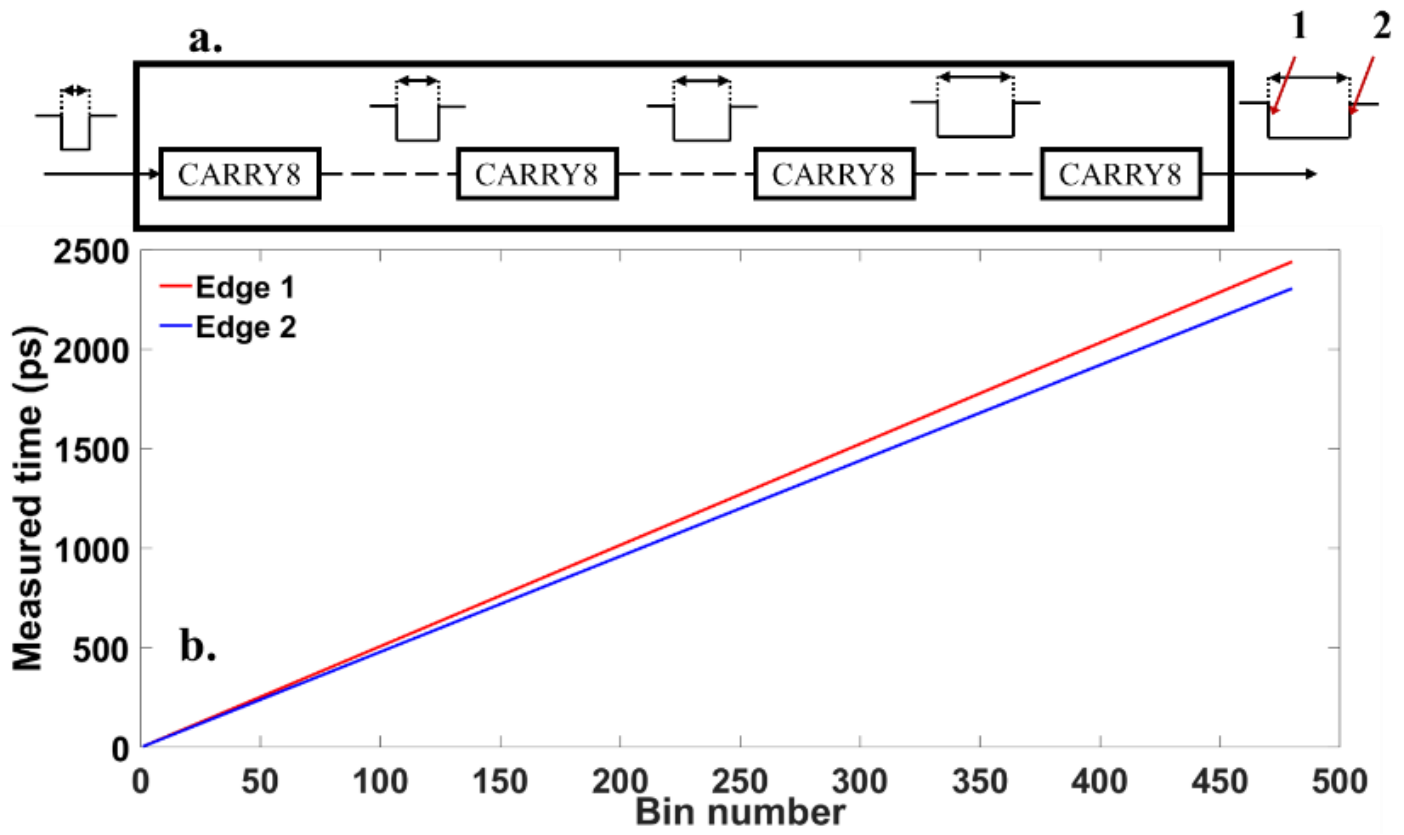

**Figure 5.** (a). An example of the propagation speed difference between the rising edge and falling edge. (b) The bin number versus the measured time curves for the rising and falling edge signals.

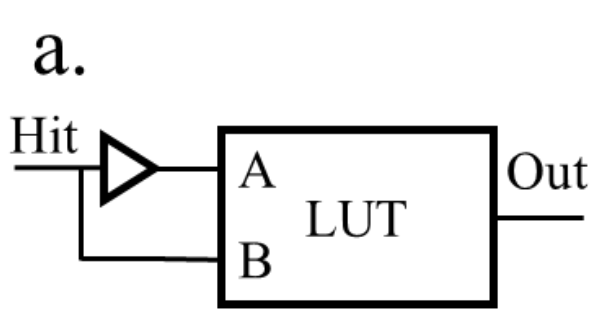

| Input | | Output |
|---|---|---|
| A | B | Out |
| 0 | 0 | 1 |
| 0 | 1 | 0 |
| 1 | 0 | 1 |
| 1 | 1 | 1 |

**Figure 6.** (a) The WU launcher built by a LUT and (b) the truth table for LUT WU launcher.

two edge signals; the edges use a different number of taps to convert a fixed time interval. Although the bin realignment proposed in [58] can remove most bubbles, it performs differently for the 0-1 and 1-0 transitions owing to the speed difference. Therefore, the bubbles cannot be removed easily in traditional WU TDCs [36]. A WU-A launcher constructed by a look-up table (LUT) was applied to our design, as shown in Figure 6(a). Figure 6(b) is the truth table for the LUT WU launcher. It generates a negative pulse when it receives a hit signal.

### *2.3. Wave Union Method Integrated With sub-TDL Structure*

The mismatch [38] in tap timing along the TDL becomes more serious when the tap interval becomes shorter and results in serious bubble problems, especially in 28 nm FPGA and more advanced process technologies [32]. Because of this more severe bubble problem and the speed difference between the rising edge and the falling edge, "The wave union method cannot be used in the UltraScale FPGA for further improving the time precision" [36]. However, we find that it is not the case in our designs.

A bubble-free method called the sub-TDL structure (or the decomposition developed independently at the same time [35]) was proposed in [55]. The main idea is to ignore the mismatch in the TDL by elongating the tap interval [55]. Song *et al.* also observed that the length of the bubble area is limited. Bubbles can be removed by decomposing the output data from the raw TDL [35].

Figure 1(b) and 1(d) show the sub-TDL structure with and without the DS method, respectively. A TDL can have several subsections with shorter thermometer codes. After encoding, the averaged TDL can be reconstructed by summing up the fine codes of the sub-TDLs. Bubbles can, therefore, be removed, and the WU method can still be applied to UltraScale FPGAs, different from the statement in [36]. The WU signal only contains two edges in this study, and the sub-TDL can be applied directly. In [63], a TDC using an 8-edge WU signal was proposed; however, the encoding process becomes much more challenging. Therefore, the sub-TDL should be modified when using more edges to improve the performance further.

### *2.4. Compensation Strategies*

To improve linearity, a fast calibration approach, called the bin compensation strategy, was demonstrated in [55]. The bin width compensation aims to compensate the bins with a negligible bin width. In this compensation strategy, two factors, the main bin calibration factor ($\mathrm{BCF_m}$, highlighted in black arrow in Figure 7) and the compensation bin calibration factor ($\mathrm{BCF_c}$, highlighted in red arrow in Figure 7), are introduced to reassign the TDC's fine codes to corrected bins. The addresses of corrected bins can be calculated based on code density tests. $\mathrm{T[k]}$ can be defined as:

$$\mathrm{T[k]} = \sum_{n=0}^{\mathrm{k}-1} W[n] = \sum_{n=0}^{\mathrm{k}-1}\{\mathrm{LSB} \times (\mathrm{DNL}[n] + 1)\}, \quad (4)$$

where $W[n]$ is the width of the *n*-th bin. $\mathrm{BCF_m}$ and $\mathrm{BCF_c}$ are calculated accordingly. For example, the uneven $\mathrm{Bin_{actual}}$ *N-2* collects a larger count proportional to its bin width in a code density test, and therefore it is necessary to assign a proportion of the count to $\mathrm{Bin_{ideal}}$ *N-1* through $\mathrm{BCF_c}$. This compensation strategy works well if the bin boundaries do not deviate too much from the ideal bin boundaries (highlighted in dash line). As each bin only has at most one $\mathrm{BCF_m}$ and one $\mathrm{BCF_c}$, this compensation method makes $\mathrm{Bin_{ideal}}$ *N+1* receiving no count allocation (due to the existence of the ultra-wide $\mathrm{Bin_{actual}}$ *N+1*, as $\mathrm{Bin_{actual}}$ *N+1* has already assigned its contribution to $\mathrm{Bin_{ideal}}$ *N+2* and *N+3*) and therefore resulting in a missing code. The compensation process can be simplified as the pseudocode below, according to [55].

```
For k = 1: N
if (T_actual [k] < T_ideal [k])
if (T_actual [k+1] < T_ideal [k])
    BCF_m = k − 1
    BCF_c = void
else if (T_actual [k+1] > T_ideal [k])
    BCF_m = k − 1
    BCF_c =k
else
    continue…
```

This method can remap and compensate bins at the same time without changing the resolution. A few missing codes can be ignored with a degraded resolution. Much severer missing code problems, however, degrade the INL and the TDC performance significantly, especially for the two-stage TDCs in [43,64]. A new compensation method still needs to be developed. We will propose a straightforward approach in this study, see Section 2.6.

Table 1 summarizes the linearity performances for the original WU design and the compensated WU design. Figure 8 presents the linearity performances of the original WU TDC and the compensated WU TDC. After compensation, small width bins have been corrected. The DNL is [-0.92, 1.75] LSB, and the INL is [-1.20, 5.97] LSB.

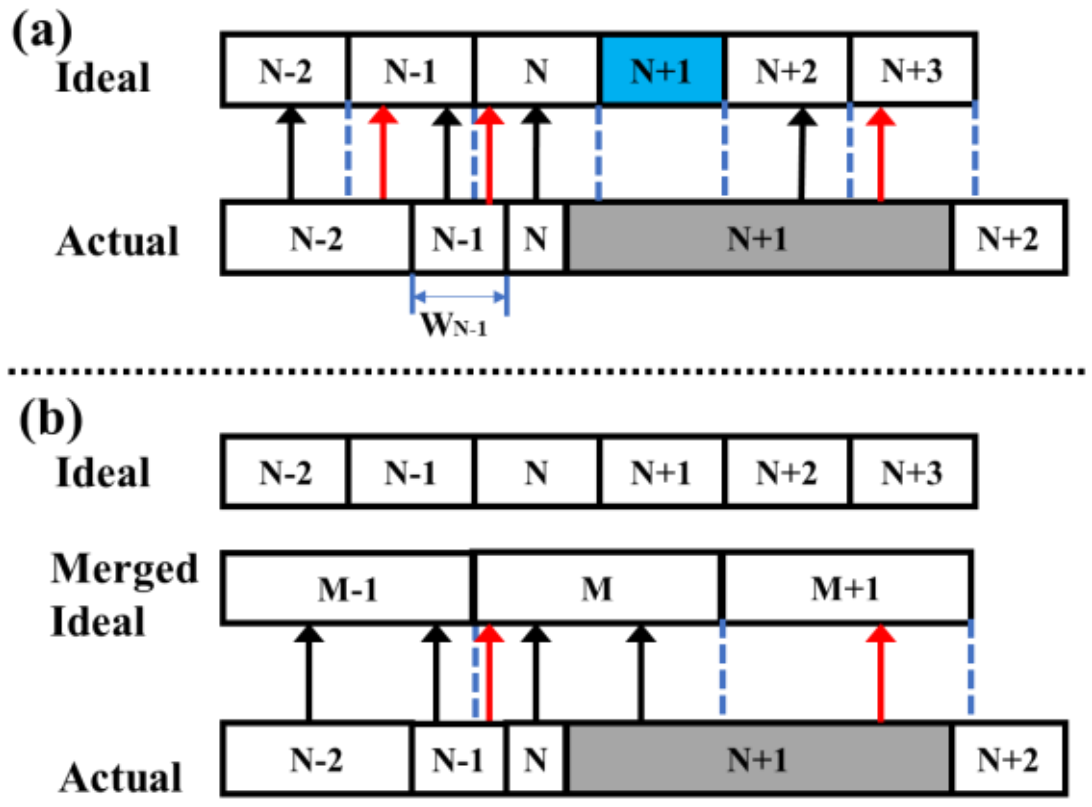

**Figure 7.** Concepts of (a) the compensation strategy and (b) the binning method.

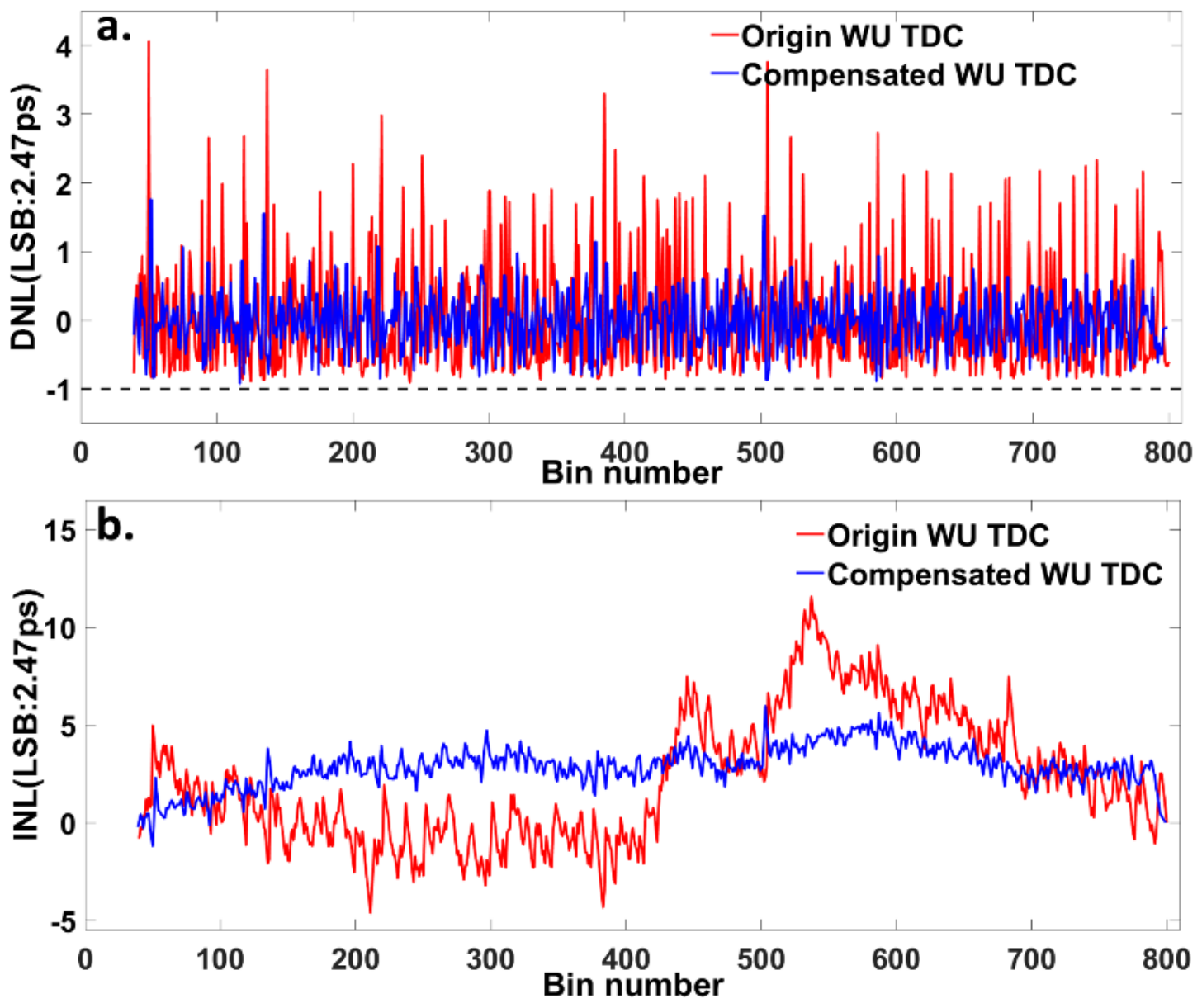

**Figure 8.** DNL and INL performances. (a) DNL and (b) INL plots of the original WU TDC and the compensated WU TDC in the UltraScale FPGA.

**Table 1.** Linearity Performance Between Original WU TDC and Compensated WU TDC

| | UltraScale (20 nm) | |
|---|---|---|
| | **Original WU TDC** | **Compensated WU TDC** |
| LSB | 2.47 ps | |
| DNL | [-0.90, 4.06] | **[-0.92, 1.75]** |
| $DNL_{pk-pk}$ | 4.96 | **2.67** |
| $\sigma_{DNL}$ | 0.82 | **0.43** |
| INL | [-4.62, 11.58] | **[-1.20, 5.97]** |
| $INL_{pk-pk}$ | 16.20 | **7.17** |
| $\sigma_{INL}$ | 3.26 | **1.06** |

### *2.5. Dual-sampling Wave Union TDC*

As shown in Figure 1 in the DS method, the CARRY8 module doubles the number of sampled taps, compared with traditional FPGA structures. As shown in Figure 4, the WU method samples the TDL taps twice, equivalently doubling sampled taps. Therefore, to achieve a better resolution, a TDC combining both, denoted as DSWU TDC, was also implemented in this study.

Figure 1(a) also shows the block diagram for this DSWU TDC, and the sub-TDL structure was adopted in this TDC. The DSWU TDC can achieve 1.23 ps resolution. The linearity, however, degrades when the resolution is improved.

*2.6. Binned Dual-sampling Wave Union TDC*

Considering the linearity performance of the DSWU TDC, a binning method can be applied to improve the linearity. It is inspired by the compensation strategy and the bin decimation method in [32]. As shown in Figure 7(b), a set of merged ideal bins are constructed by merging two consecutive bins into larger bins. Then the compensation strategy can be used to improve the linearity, even when a new bin is still small. The central concept is to reduce the numbers of ultra-wide bins and ultra-small bins, hence improving both the DNL and the INL.

## 3. Experimental Results

To evaluate the performances of the proposed TDCs, (1) code density tests and (2) time interval tests have been conducted. Two independent on-board low-jitter crystal oscillators were used as the signal sources for code density tests. There is no correlation between these two signal sources ensuring the randomness of the hit signal to the TDC clock [54]. Code density tests aim to assess the linearity performances, whereas time interval tests aim to obtain the measurement errors and the root-mean-square (RMS) resolution.

*3.1. Linearity Test Results*

Parameters like DNL, INL and standard deviations ($\sigma_{DNL}$ and $\sigma_{INL}$) are used to evaluate the linearity of a TDC. Two equations were derived to assess the equivalent bin width and its standard deviation, summarized in [65]:

$${\sigma_{eq}}^2 = \sum_{i=1}^{N}\left(\frac{W[i]^2}{12} \times \frac{W[i]}{W_{total}}\right) \text{ where } W_{total} = \sum_{i=1}^{N} W[i], \quad (5)$$

$$w_{eq} = \sigma_{eq}\sqrt{12} = \sqrt{\sum_{i=1}^{N}\left(\frac{W[i]^3}{W_{total}}\right)}, \quad (6)$$

where $w_{eq}$ is the equivalent bin width and $\sigma_{eq}$ is the equivalent bin width standard deviation, which was named differently by other groups, for example, the quantization error $\sigma_Q$ [66].

Table 1 summarizes the linearity performances of the WU TDC before and after the compensation. With the compensation strategy, the linearity of the WU TDC has been improved. The $\mathrm{DNL}_{pk-pk}$ (peak-to-peak DNL) has been enhanced from 4.96 to 2.67 LSB, whereas $\mathrm{INL}_{pk-pk}$ has also been improved significantly after the compensation.

The linearity performances of the four proposed TDCs have been summarized in Table 2. Compared with the original WU TDC and the DS TDC, the WU method is still advantageous in easing ultra-wide bin problems. The $\mathrm{DNL}_{pk-pk}$ of the original WU TDC is 4.96 LSB, but that of the DS TDC is 5.60 LSB. The compensation strategy is less suitable to be applied to the DS TDC because of the widespread presence of ultra-wide bins.

Figure 9 shows the DNL and INL curves for the DSWU TDC. The resolution of the DSWU TDC is 1.23 ps. However, the linearity degrades when the resolution is improved. The $\mathrm{DNL}_{pk-pk}$ is 8.66 LSB and the $\mathrm{INL}_{pk-pk}$ is 35.81 LSB.

The DNL and INL curves for the binned DSWU TDC are shown in Figure 10. With the binning method, the DNL is [-0.93, 1.67] LSB and the INL is [-4.13, 2.01] LSB.

The binning method is still advantageous; even the test results indicate that the compensated WU TDC and the binned DSWU TDC have similar performances in the linearity and the resolution. For the compensated WU TDC, the linearity can be improved by correcting the bins with an ultra-small width. The corrections distort the bin widths. However, the binning method can improve the linearity with less distortion by merging finer bins (LSB: 1.23 ps) into larger bins (LSB: 2.48 ps). Figure 11 presents the bin width distributions of the compensated WU TDC and the binned DSWU TDC; it shows a more concentrated bin width contribution for the binned DSWU TDC. It means the binned DSWU TDC can deliver more robust results. The standard deviations of DNL ($\sigma_{DNL}$) of the compensated WU TDC (0.43 LSB) and the binned DSWU TDC (0.35 LSB) also support this statement.

**Table 2.** Comparison of The Linearity Performances Between Four Different TDC Designs

| | UltraScale (20 nm) | | | |
|---|---|---|---|---|
| | **Compensated WU** | **DS** | **DSWU** | **Binned DSWU** |
| LSB (ps) | 2.47 | 2.53 | 1.23 | 2.48 |
| DNL | [-0.92, 1.75] | [-0.99, 4.61] | [-0.85, 7.81] | [-0.93, 1.67] |
| $\text{DNL}_{pk-pk}$ | 2.67 | 5.6 | 8.66 | 2.60 |
| $\sigma_{DNL}$ | 0.43 | 1.06 | 0.92 | 0.35 |
| INL | [-1.20, 5.97] | [-3.79, 13.69] | [-22.28, 13.53] | [-4.13, 2.01] |
| $\text{INL}_{pk-pk}$ | 7.17 | 17.48 | 35.81 | 6.14 |
| $\sigma_{INL}$ | 1.06 | 3.51 | 7.65 | 1.18 |
| $w_{eq}$ (ps) | 2.99 | 6.39 | 2.92 | 2.95 |
| $\sigma_{eq}$ (ps) | 0.86 | 1.84 | 0.84 | 0.85 |

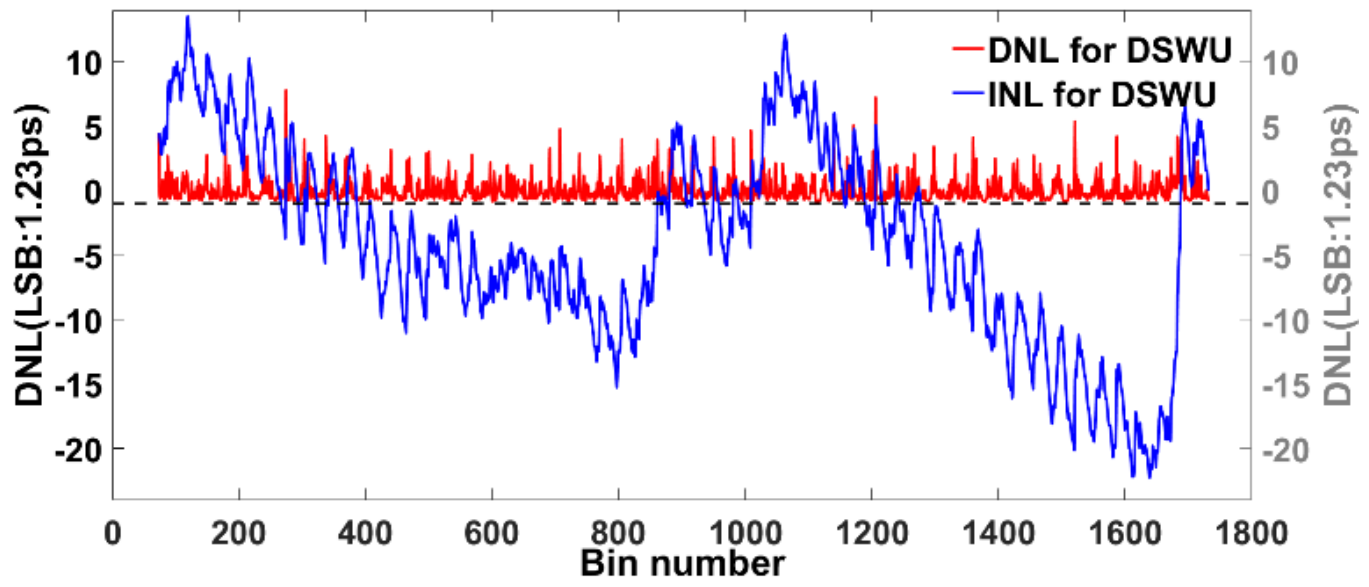

**Figure 9.** DNL and INL curves for the DSWU TDC.

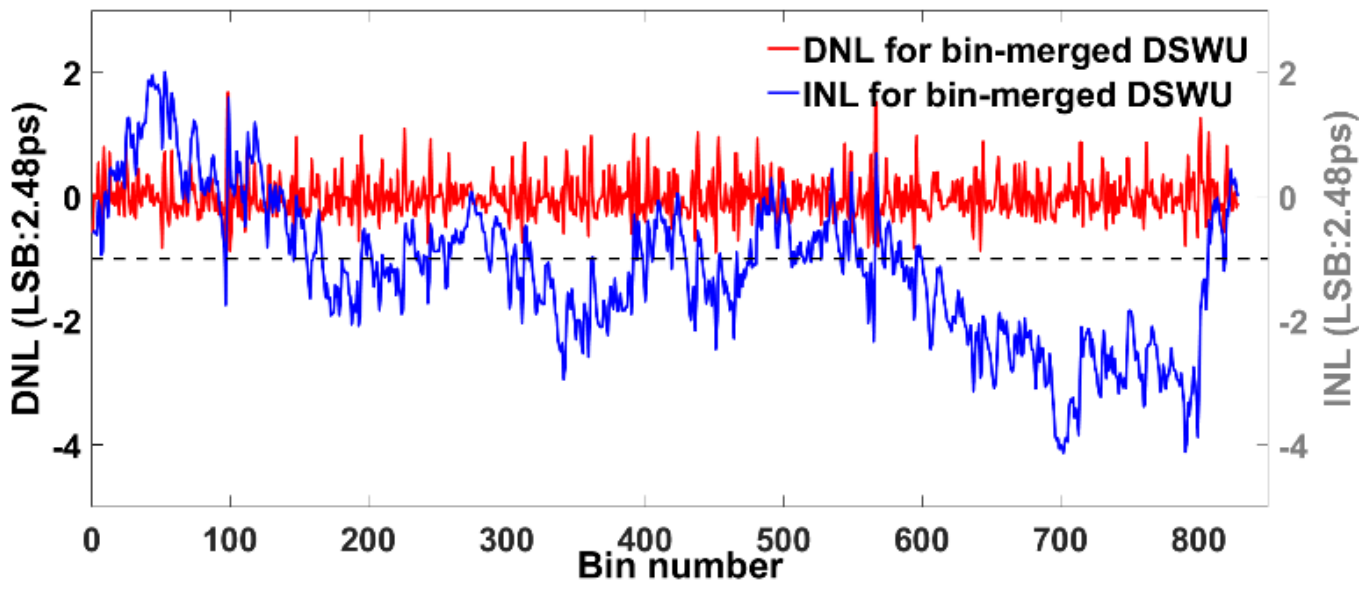

**Figure 10.** DNL and INL curves for the binned DSWU TDC.

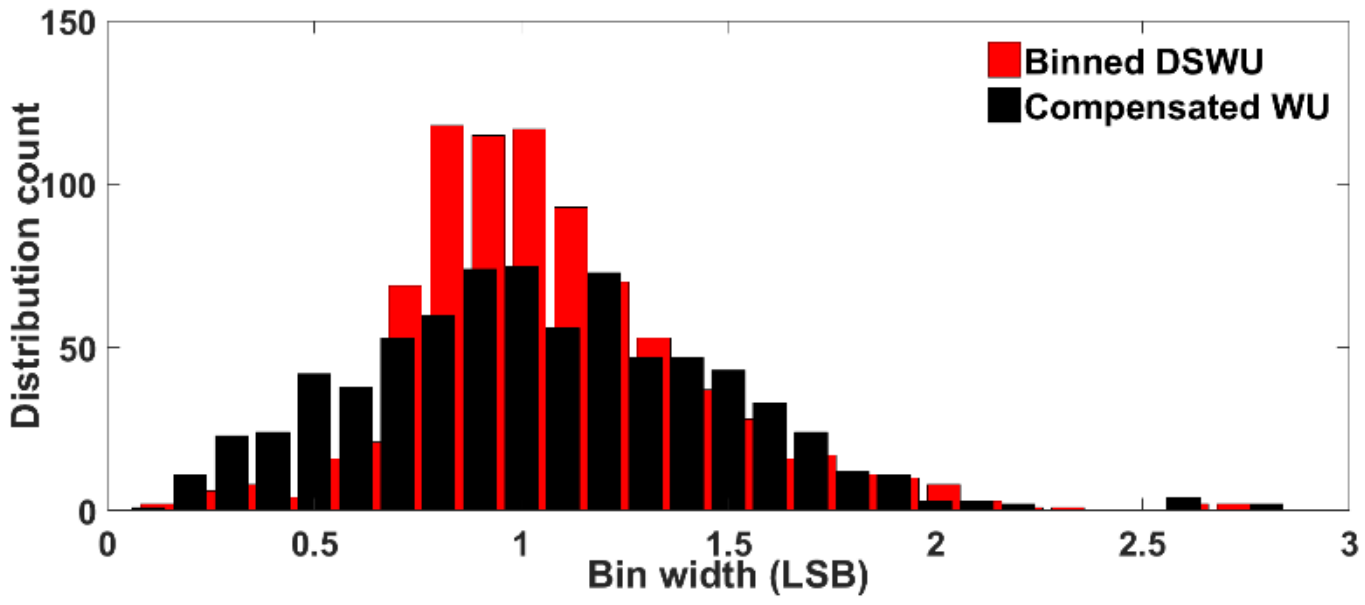

**Figure 11.** Bin width distributions for the compensated WU TDC and the binned DSWU TDC.

### 3.2. Time Interval Measurements

The precision can be estimated by the standard deviation of the distribution of repeated measurements, and it can be affected by clock jitters, jitters of input signals, electronic noise and process, voltage and temperature (PVT) variations [67]. According to [68], the precision of the whole TDC system in our case can be expressed as:

$$\sigma_{system}^2 = \sigma_{start}^2 + \sigma_{INL}^2 + \sigma_{qav}^2 + \sigma_{extra}^2, \quad (7)$$

where $\sigma_{start}$ is the start signal jitter, $\sigma_{INL}$ is the integral nonlinearity standard deviation, $\sigma_{qav}$ is the average quantization error, and $\sigma_{extra}$ represents internal jitter introduced by the FPGA device.

The delay element, IODELAYE3, was used to generate a delay with a controllable time interval between the delayed signal and the original signal. The time interval was measured by the proposed TDC and an oscilloscope, Teledyne LeCroy WaveRunner 640Z. The temperature of the testing environment was maintained, and an IDELAYCTRL module was introduced to reduce the impact of PVT variations. The step of the controllable time intervals was set to be 9.41 ps. Measurements were repeated 50,000 times for a fixed time interval.

The measurement error ($E$) is the difference between the measured value and the actual value. The measurement errors for the four TDC systems are shown in Figure 12 and Figure 13. These results were obtained from the time interval tests. The average measurement errors are around 5 ps for the proposed four TDC systems. Figure 14 contains the RMS resolutions for four TDC systems with the jitters from IODELAYE3 also included. The RMS resolution for the DS TDC system is 5.74 ps with $\sigma_{RMS} = 1.09$ ps. The RMS resolution for the WU TDC system is 5.95 ps with $\sigma_{RMS} = 0.77$ ps. The DSWU TDC system achieves an RMS resolution of 5.69 ps with $\sigma_{RMS} = 1.05$ ps. The binned DSWU TDC system reaches an RMS resolution of 5.89 ps with $\sigma_{RMS} = 0.46$ ps.

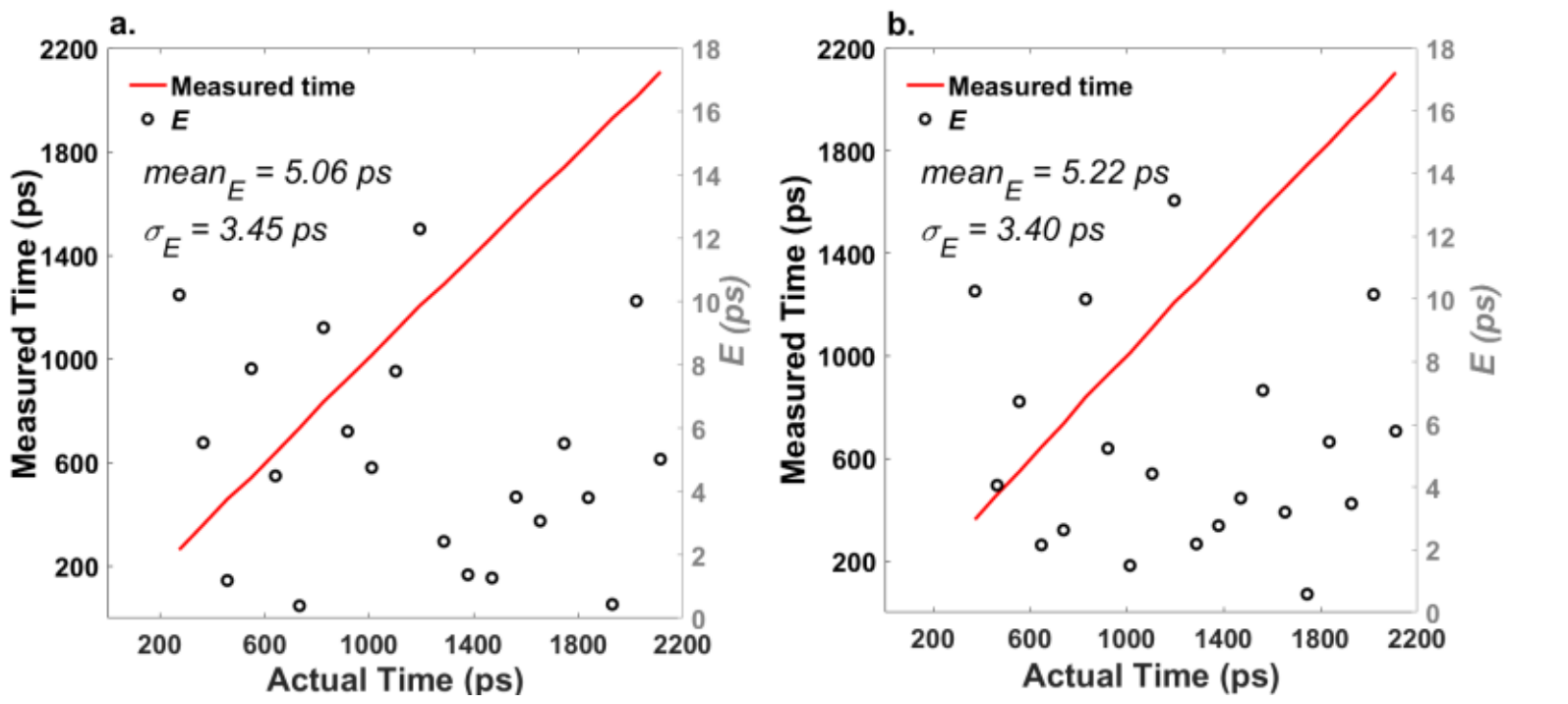

**Figure 12.** Measurement errors obtained from the time interval tests for (a) the WU TDC system and (b) the DS TDC system.

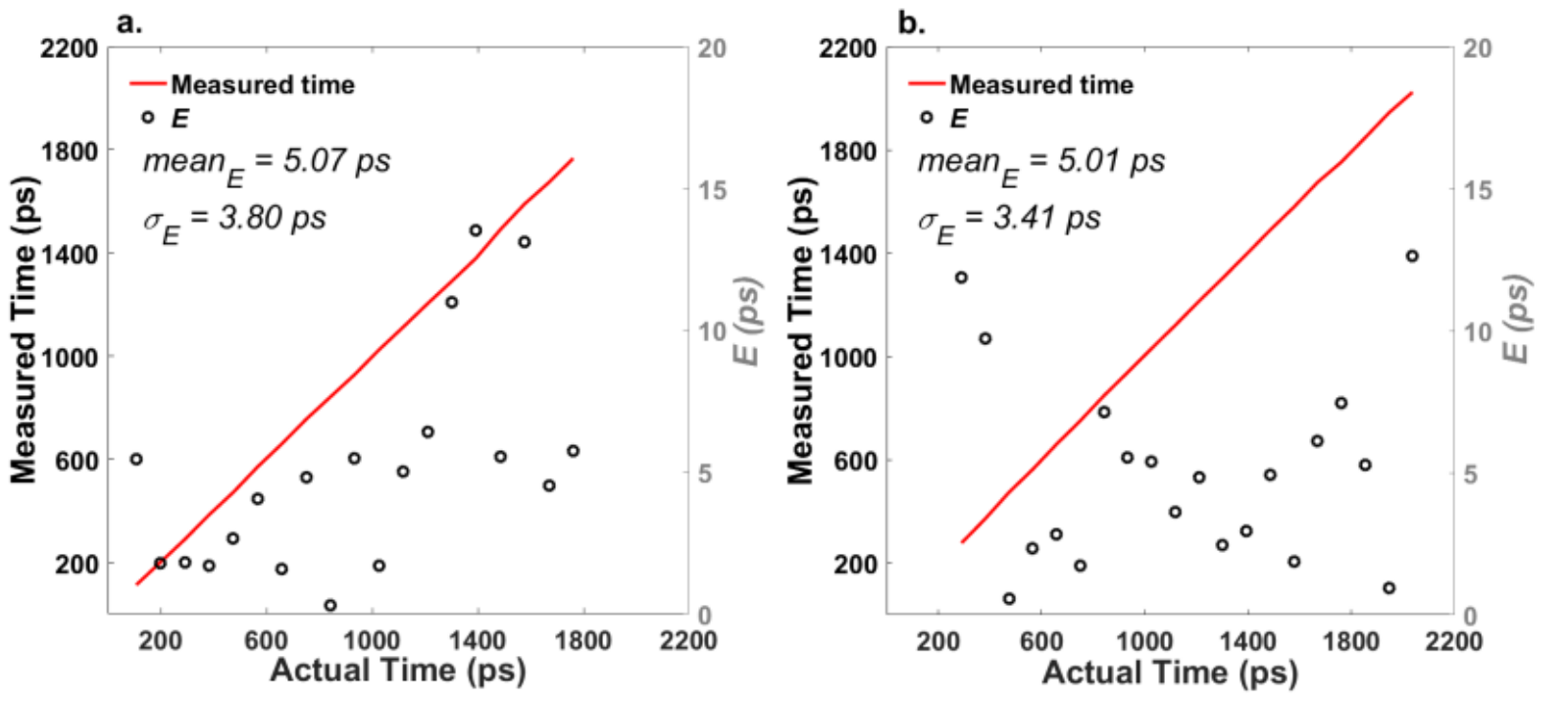

**Figure 13.** Measurement errors obtained from the time interval tests for (a) the DSWU TDC system and (b) the binned DSWU TDC system.

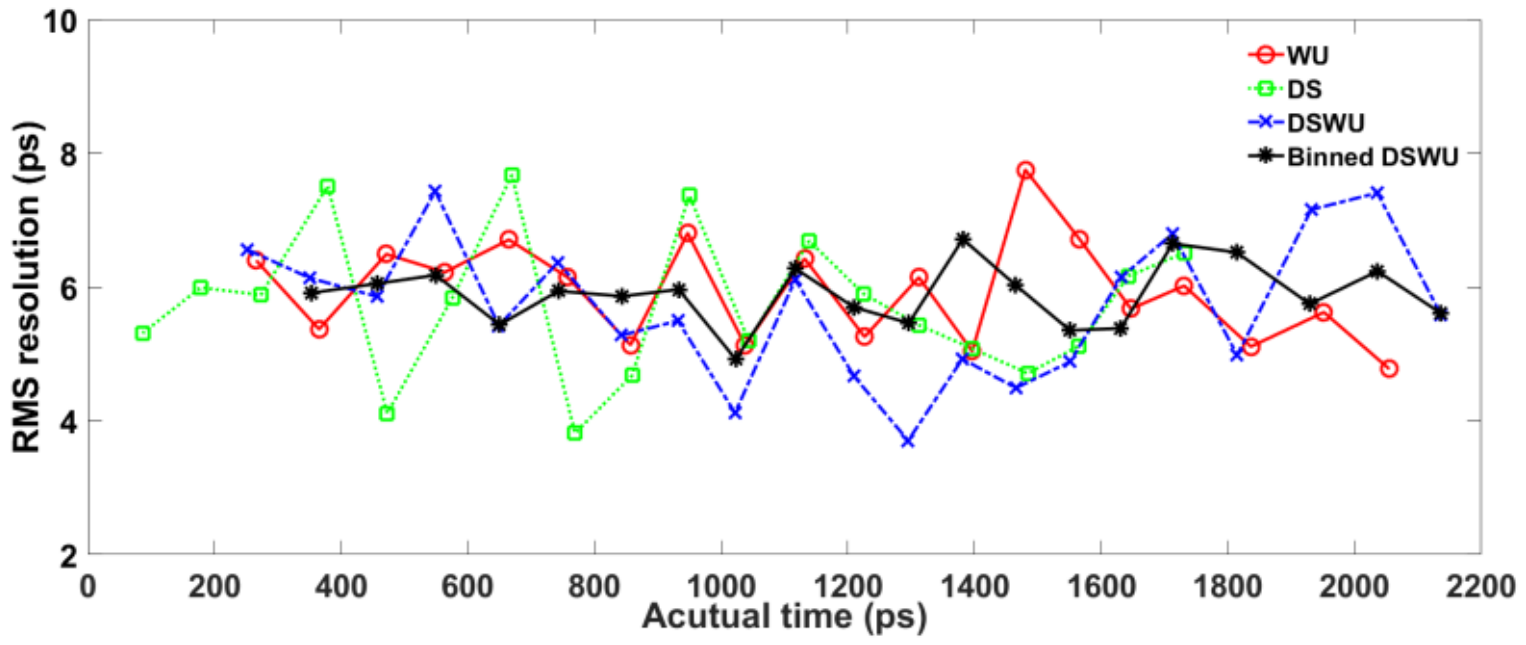

**Figure 14.** RMS resolutions obtained by the time interval tests for four TDC systems (jitters from IODELAY3 and the clock signal are included).

### 3.3. Analysis of Measurement Uncertainties

Szplet *et al.* proposed another way to estimate measurement uncertainties by analyzing error sources for two-stage multi-phase TDC [68]. Their analysis approach can be extended for our single-stage TDCs with or without the wave union launcher, and the RMS resolution can be expressed as:

$$\sigma_{system}^2 = \sigma_{sig}^2 + \sigma_{eq}^2 + \sigma_{clk}^2, \tag{8}$$

In Equation (8), $\sigma_{sig}$ is denoted as the total jitter when a signal transmits through the delay lines and can be calculated as:

$$\sigma_{sig}^2 = \sigma_{wu}^2 + \sigma_{DL}^2, \tag{9}$$

where $\sigma_{wu}$ is the jitter caused by the wave union launcher and several delay elements in the tapped delay line ($\sigma_{DL}$) are involved. For the wave union launcher, it is constructed by a LUT and a CARRY8. And in each CARRY8, there are eight delay elements ($\sigma_{CY}$). Therefore, $\sigma_{wu}$ can be expressed:

$$\sigma_{wu}^2 = 8\sigma_{CY}^2 + \sigma_{LUT}^2, \tag{10}$$

It is difficult to predict how many delay elements of the TDL are involved in measurements. However, according to [68], for a TDL with $n$ delay elements, $\sigma_{DL}$ can be expressed by its expected value:

$$\begin{aligned} \mathrm{E}[\sigma_{DL}^2] &= \sum_{i=1}^{n} \sigma_{DLi}^2 \cdot p(\sigma_{DLi}^2) = \sum_{i=1}^{n} i \cdot \sigma_{CY}^2 \cdot p(\sigma_{CY}^2) \\ &= \sum_{i=1}^{n} i \cdot \sigma_{CY}^2/n = (\sigma_{CY}^2/n) \sum_{i=1}^{n} i \\ &= \frac{n+1}{2}\sigma_{CY}^2 \approx \frac{n}{2}\sigma_{CY}^2, \end{aligned} \tag{11}$$

where $\sigma_{DL_i}$ is the jitter caused by $i$-th delay element and the $i$-th delay element has the probability $p(\sigma_{DLi}^2)$ of hitting during a signal measurement. Because the delay element used in our cases is CARRY8, $\sigma_{DL_i}^2 = i\sigma_{CY}^2$ and $p(\sigma_{DLi}^2) = p(\sigma_{CY}^2)$. From Equations (8)–(11), the RMS resolution can be expressed as:

$$\sigma_{system}^2 = \sigma_{clk}^2 + \sigma_{eq}^2 + \sigma_{LUT}^2 + (\tfrac{n}{2} + 8)\sigma_{CY}^2, \tag{12}$$

To obtain an estimated value of $\sigma_{LUT}$ and $\sigma_{CY}$, a ring oscillator (RO) has been constructed, as shown in Figure 15. There are $m$ delay elements in this RO, and the jitter in this RO can be expressed as:

$$\sigma_{RO}^2 = \sigma_{LUT}^2 + m\sigma_{CY}^2, \quad (13)$$

The measured results, as shown in Figure 16, show that $\sigma_{CY} = 0.16$ ps and $\sigma_{LUT} = 1.45$ ps. The four proposed TDCs share similar architectures, and there is no wave union launcher in the DS TDC, so $\sigma_{sig} = 2.91$ ps (for WU, DSWU and binned DSWU TDCs) and $\sigma_{sig} = 2.51$ ps (for DS TDC).

The clock signal was generated by a clock generator (Si5335A, Silicon Labs) and an MMCM (Mixed-mode Clock Manager) module. Measured by the oscilloscope, the jitter caused by the clock signal ($\sigma_{clk}$) is 4.42 ps. Considering the relatively large clock jitter, we introduced $\sigma_{TDC}$ to evaluate the RMS resolution of the TDC and it can be expressed as Equation (14). And the measurement uncertainties of four proposed TDCs have been summarized in Table 3. Table 3 also compares our analysis with two previously published works in 45 nm and 20 nm FPGAs with our analysis providing more detailed contributions from the error sources.

$$\sigma_{TDC}^2 = \sigma_{sig}^2 + \sigma_{eq}^2, \quad (14)$$

**Table 3**. Evaluation of Measurement Uncertainties

| | **Spartan-6 (45 nm)** | **UltraScale (20 nm)** | | | | |
|---|---|---|---|---|---|---|
| | [68]-2019 | [55]-2019 | Compensated WU | DS | DSWU | Binned DSWU |
| LSB (ps) | - | 5.02 | 2.47 | 2.53 | 1.23 | 2.48 |
| **Error Source Analysis** | | | | | | |
| $\sigma_{clk}$ (ps) | 1.93 | - | 4.42 | | | |
| $\sigma_{CY}$ (ps) | 0.153 | - | 0.16 | | | |
| $\sigma_{LUT}$ (ps) | 1.33 | - | 1.45 | | | |
| $\sigma_{sig}$ (ps) | - | - | 2.91 | 2.51 | 2.91 | 2.91 |
| $\sigma_{eq}$ (ps) | - | 1.45 | 0.86 | 1.84 | 0.84 | 0.85 |
| $\boldsymbol{\sigma_{TDC}}$ (ps) | - | - | 3.03 | 3.11 | 3.02 | 3.03 |
| $\sigma_{system}$ (ps) | 10.19 | - | 5.36 | 5.41 | 5.36 | 5.36 |
| Time Interval Test | | | | | | |
| $\sigma_{system}$ (ps) | - | 7.8 | 5.95 | 5.74 | 5.69 | 5.89 |

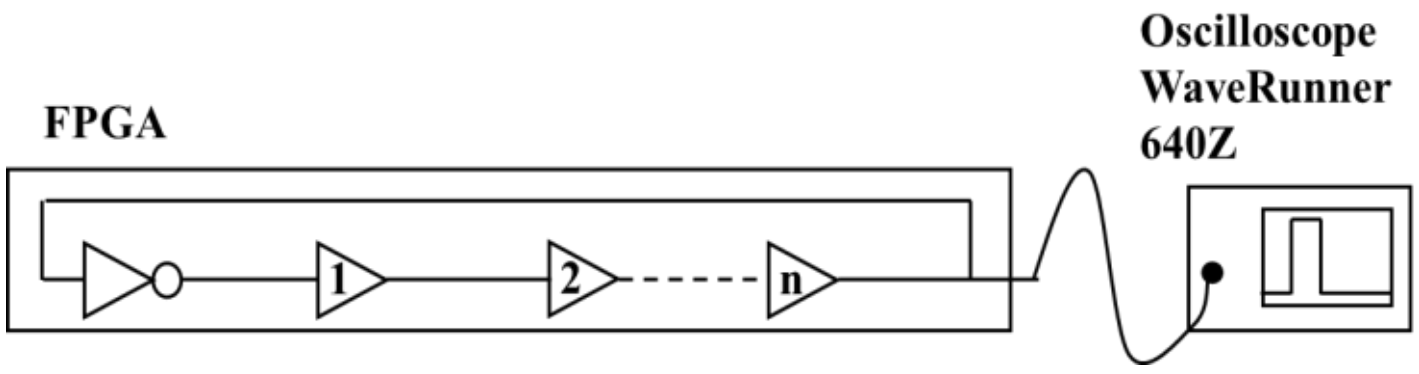

**Figure 15.** Test setup for investigating jitters of LUT and the delay element.

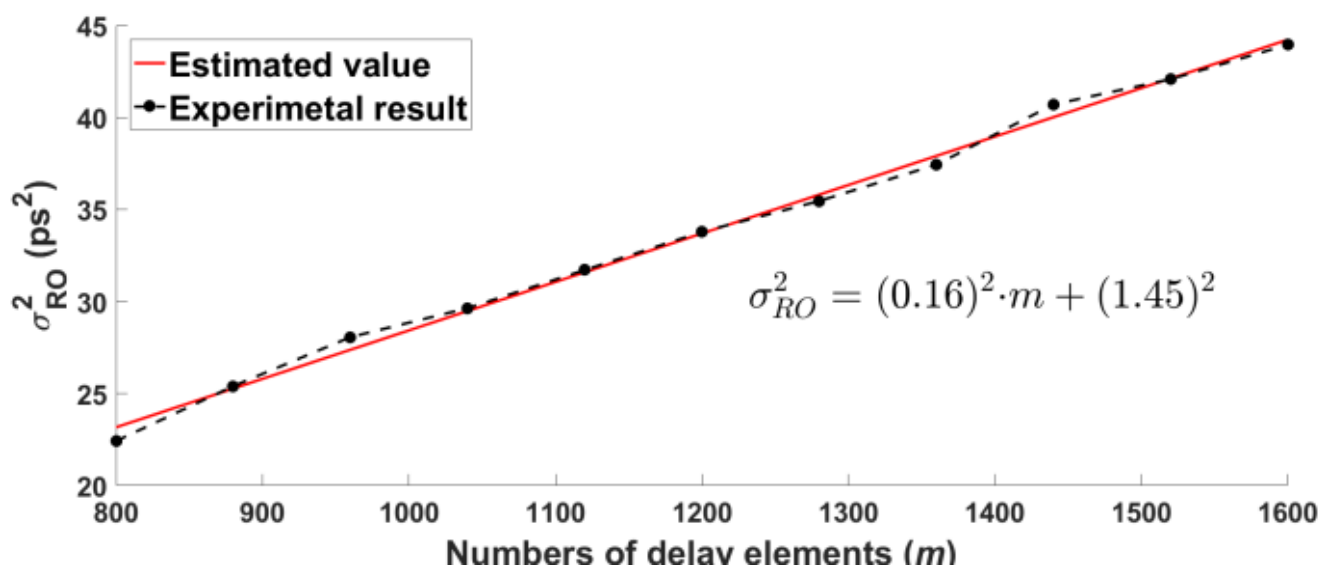

**Figure 16.** Test results for investigating jitters of LUT and the delay element.

**Table 4.** Consumption of Logic Resources

| | | w/o WU | WU | DS | DSWU/ Binned DSWU |
|---|---|---|---|---|---|
| Modules | Total | Used | Used | Used | Used |
| CARRY8 | 30300 | 80 (0.26%) | 85 (0.28%) | 76 (0.25%) | 88 (0.29%) |
| LUTs | 242400 | 703 (0.29%) | 1349 (0.56%) | 1272 (0.52%) | 2460 (1.01%) |
| FFs | 484800 | 1195 (0.24%) | 1840 (0.38%) | 2190 (0.45%) | 3463 (0.71%) |
| BRAM | 600 | 1.5 (0.25%) | 4.5 (0.75%) | 3.5 (0.58%) | 7.5 (1.25%) |
| CLB | 30300 | 271(0.89%) | 359(1.18%) | 283(0.94%) | 529(1.75%) |

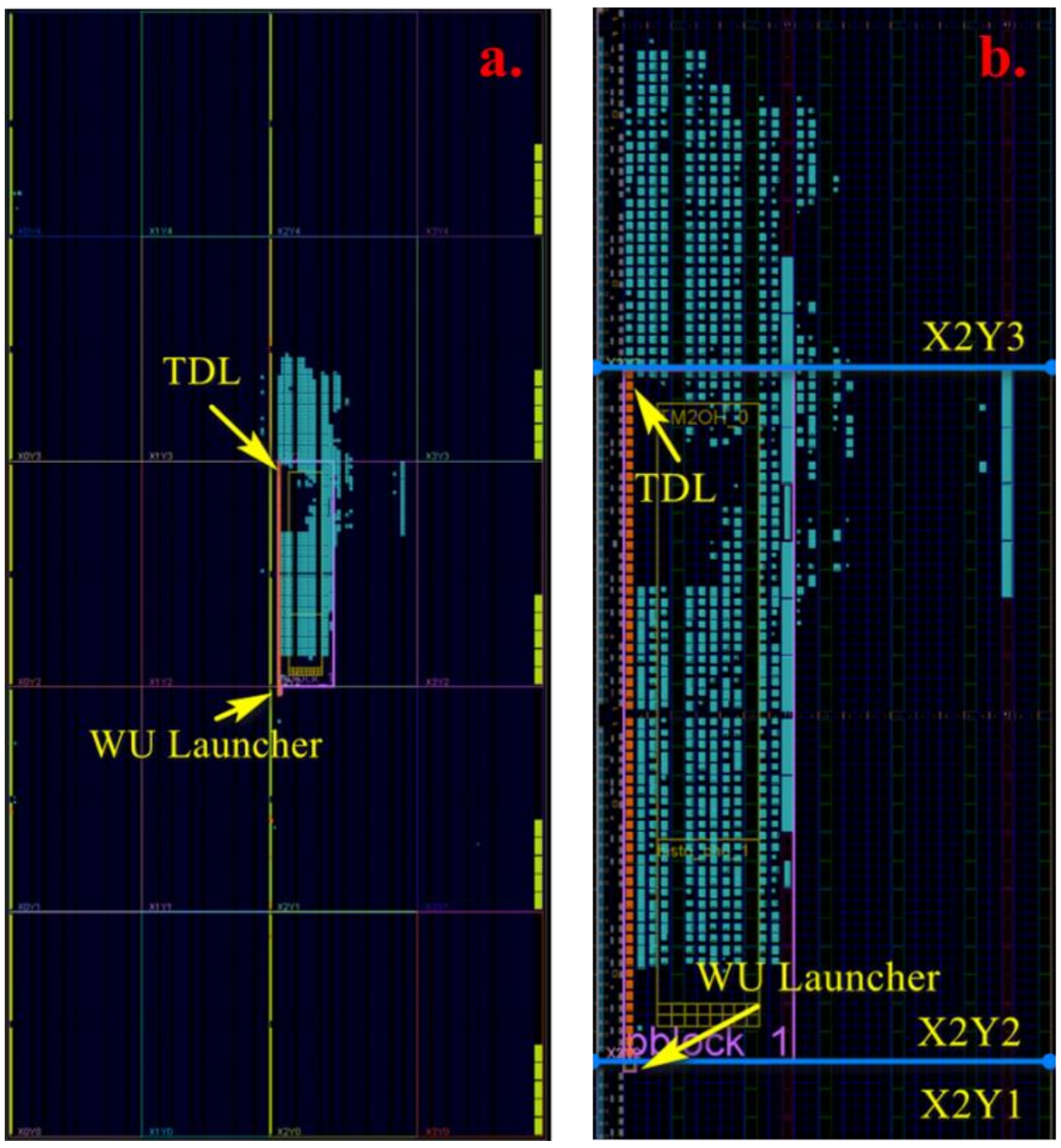

**Figure 17.** Implementation layouts of the DSWU TDC. a) Overview. b) Clock regions (X2Y1 ~ X2Y3)

### *3.4. Resource Consumption*

The consumptions of logic resources for the proposed TDCs were calculated by the EDA tools and are shown in Table 4. Extra logic resources were used when using the WU method and the DS structure (more block RAM needed to perform the histogram functions). However, the usages of logic resources for proposed TDCs are still low, showing great potential for multiple-channel applications.

The implementation layouts of the DSWU TDC are shown in Figure 17. The overview is shown in Figure 17(a), and the layout of clock regions (X2Y1~X2Y3) is shown in Figure 17(b). The layouts were obtained from the Vivado suite. To reduce the large nonlinearity contributed by the clock tree distribution, the TDL is confined in 60 logic cells within a central clock region (Slice X49Y120~X49Y179). The WU launcher is confined in Slice X49Y119. These constraints are also applicable to the WU TDC and the binned DSWU TDC. For the DS TDC, the constraints for the WU launcher are not required. Figure 17 shows that the layout does not consume much hardware resources, and therefore the proposed TDCs are suitable for multichannel applications. If a longer measurement range is required, coarse counters can be easily included for the proposed TDCs.

**Table 5.** Comparison of Published FPGA-based TDCs and the Proposed TDCs

| Ref - Year | Methods | Device | LSB (ps) | RMS Resol. (ps) | DNL, $DNL_{pk-pk}$ (LSB) | INL, $INL_{pk-pk}$ (LSB) | Power (mW) |
|---|---|---|---|---|---|---|---|
| [69]-13* | Multichannel, TDL | Virtex-5 | 30.00 | 15.00 | [-1.00, 3.00], 4.00 | [-4.00, 4.00], 8.00 | N/S |
| [70]-13* | Multichannel, TDL | Virtex-6 | 10.00 | 18.50 | [-1.00, 1.50], 2.50 | [-2.25, 1.61], 3.86 | N/S |
| [71]-14* | Multichannel, TDL | Kintex-7 | 22.70 | N/S | N/S, 2.60 | N/S, 3.40 | N/S |
| [53]-14* | WU-A, TDL | Cyclone II | 20.00 | 21.00 | N/S | N/S | N/S |
| [58]-15* | WU-A, multichannel, TDL | Kintex-7 | N/S | < 10.00 | N/S | N/S | N/S |
| [39]-15* | Multichain averaging, TDL | Virtex-6 | 1.50 M=16 | 4.20 M=16 | [-0.70, 0.80], 1.50 (M=8, LSB=24ps) | [-1.00, 0.70], 1.70 (M=8, LSB=24ps) | N/S |
| [44]-16* | Dual-phase, TDL | Virtex-6 | 10.00 | 12.80 | [-1.00, 1.91], 2.91 | [-2.20, 3.93], 6.13 | N/S |
| [72]-16* | WU-A, multichain, 2-stage interpolation, TDL | Spartan-6 | 0.90 | < 6.00 | [-1.00, 6.25] [1] | [-26.20, 11.50], 37.70 | N/S |
| [34]-16* | Tuning-TDL, heterogenous sampling | Kintex-7<br>Virtex-6<br>Spartan-6 | 10.6<br>10.1<br>16.7 | N/S<br>N/S<br>N/S | [-1.00, 1.45], 2.45<br>[-1.00, 1.18], 2.18<br>[-1.00, 1.22], 2.22 | [-1.23, 4.30], 5.53<br>[-3.03, 2.46], 5.49<br>[-0.70, 2.56], 3.26 | N/S |
| [36]-16* | Dual sampling | Ultra-Scale | 2.25 | 3.90 | [-1.00, 4.78] [1] | N/S | N/S |
| [73]-17* | Tuned-TDL | Virtex-7 | 10.5 | N/S | [-0.38, 0.87], 1.25 | [-1.23, 1.02], 2.25 | N/S |
| [74]-17* | Multichain, TDL | Virtex-7 | 1.15 | 3.50 | [-0.98, 3.50], 4.48 | [-5.90, 3.10], 9.00 | N/S |
| [45]-18* | Multi-phase | Cyclone V | 1.56 | 2.30 | [-1.00, 5.60] [1] | [-8.00. 35.00] [1] | N/S |
| [55]-19* | Multichannel, sub-TDL | Ultra-Scale | 5.02 | 7.80 | [-0.12, 0.11], 0.27 | [-0.18, 0.46], 0.59 | N/S |
| [75]-19* | Super-WU, multichannel | Artix-7 | 2.00 | < 12.50 | N/S | N/S, 2.10 | < 10 |
| [63]-19* | WU-A, TDL | Kintex-7 | 1.77 | 3.00 | [-1.00, 4.50] [1] | [-37.70, 12.00] [1] | < 1.02 |

**Table 5.** (*Continued.*) Comparison of Published FPGA-based TDCs and the Proposed TDCs

| | | | | | | | |
|---|---|---|---|---|---|---|---|
| **This Work▲** | **1) Sub-TDL, compensation, WU-A** | Ultra-Scale | **2.47** | 5.95 [2] **3.03** [3] | **[-0.92, 1.75], 2.67** | **[-1.20, 5.97], 7.17** | **0.92** |
| | **2) Sub-TDL, dual sampling** | | **2.53** | 5.74 [2], **3.11** [3] | **[-0.99, 4.61], 5.60** | **[-3.79, 13.69], 17.48** | **0.88** |
| | **3) Sub-TDL, dual sampling, WU-A** | | **1.23** | 5.69 [2], **3.02** [3] | **[-0.85, 7.81], 8.66** | **[-22.28, 13.53], 35.81** | **1.03** |
| | **4) Sub-TDL, dual sampling, WU-A, binning** | | **2.48** | 5.89 [2], **3.03** [3] | **[-0.93, 1.67], 2.60** | **[-4.13, 2.01], 6.14** | **1.03** |

[1] Approximate values from the figures; [2] Data obtained from time interval tests; [3] Data obtained from the analysis of measurement uncertainties; * Modelling method unknown; ▲ Modelled with Verilog HDL.

*3.5. Discussions*

We used Verilog hardware description language to model the proposed TDCs. VHDL can also be used to build a TDC. However, most TDCs are built with cell primitives provided by FPGA manufactures [30]. The cell primitives are intrinsic to the target architecture and cannot be changed by the modelling methods and EDA tools [76,77]. With cell primitives, redundant logic resource consumption can be avoided.

Table 5 summarizes the key parameters of the proposed TDCs against other recently proposed advanced TDCs. The power consumptions of the proposed TDCs were estimated by Vivado Design Suite, and the results show that the proposed TDCs have comparable performances in power consumption, compared with some earlier TDCs [63,75]. Thanks to recent advances in CMOS manufacturing technologies and research in TDC architectures, the resolution (LSB) has been significantly improved from 30.00 ps to 0.90 ps. However, there are trade-offs between the resolution and the linearity. The improvements in the linearity are not usually at the same pace as those in the resolution.

Compared with previously published works (the 16-chain TDC design in [39]), the LSB of the DSWU TDC has been significantly improved with acceptable consumption of logic resources. The DSWU TDC also performs better in the linearity and the logic consumption than other previously published TDCs [45,72,74] with a close or similar LSB. In [45,72,74] multi-phase or multi-TDL channels were used, whereas, in our methods, only a single TDL channel was used. Compared with the previously published DS TDC in [36], the sub-TDL structure can remove zero-width bins (see [36] and this work in Table 5) and this feature can be further exploited to improve the TDC performance. Zero-width bins can be ignored when processing the TDC outputs in the TDC in [36], but this degrades the resolution of TDC.

The combination of the WU method with the sub-TDL structure also shows promising results, suitable for multichannel applications.

Experimental results show that the binning method seems an applicable method to improve the linearity and in our study the $\mathrm{DNL}_{pk-pk}$ has been enhanced to 2.60 LSB and the $\mathrm{INL}_{pk-pk}$ has been improved to 6.14 LSB. If the merged bins are wide enough, a long TDL can be constructed regardless of the large clock skews. Note that the binning method shows great potential for some LiDAR systems that require an acceptable resolution (~50 ps) with high linearity [78].

The estimation of precision becomes a challenging issue when the resolution is close to 1 ps. The experimental RMS resolution can be affected by the experimental devices, such as the clock generator and the oscilloscope. The precision can also be obtained by analyzing the error sources. For a TDC design, once the general architecture ($\sigma_{sig}$) has been fixed, the equivalent standard deviation of the bin widths ($\sigma_{eq}$ or the quantization error $\sigma_Q$ defined in [66]) will be the only factor that varies when different methods are implemented. From Table 3, both the WU TDC and DSWU TDC can reduce $\sigma_{eq}$ by improving the resolution. For the binned DSWU TDC, $\sigma_{eq}$ is still low, because its linearity has been improved significantly with a loss in the resolution (LSB: from 1.23 ps to 2.48 ps). Therefore, these TDCs achieve similar precision. In our study, the measurement uncertainties of the four proposed TDCs ($\sigma_{TDC}$: ~3 ps) are mainly contributed by architecture-dependent jitters $\sigma_{sig}$ ($\sigma_{sig}$ = 2.51 ps for the DS without the wave union launcher and $\sigma_{sig}$ = 2.91 ps for the WU, DSWU and binned DSWU TDCs). By analyzing the error sources, we can conclude that the precisions of the proposed WU-related TDCs ($\sigma_{eq}$ ~ 0.85 ps) reach the upper limit of their architectures, and the precision of DS TDC, however, can still be further improved with $\sigma_{eq}$ = 1.84 ps.

From Table 3, $\sigma_{system}$ obtained from time interval tests is slightly larger than that obtained through the detailed analysis of measurement uncertainties. It indicates that the IODELAY module could introduce extra jitter to the tested systems.

The potential for a multichannel TDC design is essential for broader applications. Compared with recently published works, the proposed TDCs are efficient in the consumption of logic resources. A 7.4 ps TDC was proposed in [79] claiming to consume low logic resources. Their solution, however, was not without a price, as the routing resources were used as the delay elements. Even the consumption of logic elements was low; it still needed to use a large logic area to constraint the routing paths: 1024 configurable logic blocks (CLBs) [79]. In comparison with this design, the proposed TDCs use a smaller logic area (shown in Table 4). Therefore, as high-performance TDCs in UltraScale devices, the proposed TDCs show great potential for multichannel applications due to the low logic resource consumption.

## 4. Conclusion

In this paper, we proposed and evaluated:

1) Four TDC architectures by integrating the WU method and the DS structure with the sub-TDL topology;
2) The first efficient WU TDC in UltraScale FPGA (Different from the previously reported study [36], we concluded that the WU method is still useful in UltraScale FPGAs when it integrates sub-TDL structures.);
3) A binning method to improve the linearity;
4) Detailed analysis of measurement uncertainties and error sources for the four proposed TDCs.

With the sub-TDL architecture, bubble problems have been significantly alleviated. Meanwhile, the WU method can still be applied to improve the resolution in UltraScale FPGA devices. The compensated method has been used to correct the conversion bias and the deviation of the bin width directly with acceptable logic resources. The DS structure is another way to improve the resolution, although it does not perform better in comparison with the WU method.

By integrating these methods, a TDC with 1.23 ps resolution was also implemented and evaluated. A binning method was implemented to the proposed DSWU TDC to improve the linearity. The binning method can also be implemented with multi-chain design strategies to improve the linearity with an acceptable resolution and be preferable to some LiDAR systems. Latest TDC architectures show limited precision due to device mismatch problems prevailed in advanced manufacturing technologies. For our proposed TDCs, when the tapped delay line method is fixed, the precision can be improved by using a better clock source and by reducing $\sigma_{eq}$. However, the accumulated jitters caused by delay elements will mainly contribute to the precision of a TDC.

Our solutions demonstrate comparable improvements in the resolution compared with previously published works, listed in Table 5. Moreover, compared with other published TDCs with high resolution (~1 ps) in [45,72,74], the proposed 1.23 ps resolution TDC requires less hardware. In [45,72,74], the multi-phase or multi-TDL channel was used, whereas in this report, only a single TDL channel was used. With low consumptions of logic resources, the proposed TDCs also show great potential for multichannel applications.

**Funding:** The research has been supported by the Engineering and Physical Sciences Research Council under EPSRC Grant: EP/M506643/1 and the Royal Society of Edinburgh.

**Acknowledgements:** We would like to acknowledge the support from Xilinx for donating UltraScale FPGA development kits to the research group.

**Conflicts of Interest:** The authors declare no conflict of interest.

## References

1. Niclass, C.; Favi, C.; Kluter, T.; Gersbach, M.; Charbon, E. A 128 x 128 Single-Photon Image Sensor With Column-Level 10-Bit Time-to-Digital Converter Array. *IEEE J. Solid-State Circuits* **2008**, *43*, 2977–2989, doi:10.1109/JSSC.2008.2006445.
2. Gersbach, M.; Maruyama, Y.; Trimananda, R.; Fishburn, M.W.; Stoppa, D.; Richardson, J.A.; Walker, R.; Henderson, R.; Charbon, E. A Time-Resolved, Low-Noise Single-Photon Image Sensor Fabricated in Deep-Submicron CMOS Technology. *IEEE J. Solid-State Circuits* **2012**, *47*, 1394–1407, doi:10.1109/JSSC.2012.2188466.

3. Nolet, F.; Lemaire, W.; Dubois, F.; Roy, N.; Carrier, S.; Samson, A.; Charlebois, S.A.; Fontaine, R.; Pratte, J.-F. A 256 Pixelated SPAD readout ASIC with in-Pixel TDC and embedded digital signal processing for uniformity and skew correction. *Nucl. Instrum. Methods Phys. Res. Sect. Accel. Spectrometers Detect. Assoc. Equip.* **2020**, *949*, 162891, doi:10.1016/j.nima.2019.162891.
4. Lecoq, P. Pushing the Limits in Time-of-Flight PET Imaging. *IEEE Trans. Radiat. Plasma Med. Sci.* **2017**, *1*, 473–485, doi:10.1109/TRPMS.2017.2756674.
5. Gariepy, G.; Tonolini, F.; Henderson, R.; Leach, J.; Faccio, D. Detection and tracking of moving objects hidden from view. *Nat. Photonics* **2016**, *10*, 23–26, doi:10.1038/nphoton.2015.234.
6. Tang, Y.; Hu, Y.; Cui, J.; Liao, F.; Lao, M.; Lin, F.; Teo, R.S.H. Vision-Aided Multi-UAV Autonomous Flocking in GPS-Denied Environment. *IEEE Trans. Ind. Electron.* **2019**, *66*, 616–626, doi:10.1109/TIE.2018.2824766.
7. Song, H.; Choi, W.; Kim, H. Robust Vision-Based Relative-Localization Approach Using an RGB-Depth Camera and LiDAR Sensor Fusion. *IEEE Trans. Ind. Electron.* **2016**, *63*, 3725–3736, doi:10.1109/TIE.2016.2521346.
8. Li, X.; Yang, B.; Xie, X.; Li, D.; Xu, L. Influence of Waveform Characteristics on LiDAR Ranging Accuracy and Precision. *Sensors* **2018**, *18*, 1156, doi:10.3390/s18041156.
9. Li, D.D.-U.; Arlt, J.; Tyndall, D.; Walker, R.; Richardson, J.; Stoppa, D.; Charbon, E.; Henderson, R.K. Video-rate fluorescence lifetime imaging camera with CMOS single-photon avalanche diode arrays and high-speed imaging algorithm. *J. Biomed. Opt.* **2011**, *16*, 096012, doi:10.1117/1.3625288.
10. Henderson, R.K.; Johnston, N.; Mattioli Della Rocca, F.; Chen, H.; Day-Uei Li, D.; Hungerford, G.; Hirsch, R.; Mcloskey, D.; Yip, P.; Birch, D.J.S. A 192x128 Time Correlated SPAD Image Sensor in 40-nm CMOS Technology. *IEEE J. Solid-State Circuits* **2019**, *54*, 1907–1916, doi:10.1109/JSSC.2019.2905163.
11. Li, D.D.-U.; Ameer-Beg, S.; Arlt, J.; Tyndall, D.; Walker, R.; Matthews, D.R.; Visitkul, V.; Richardson, J.; Henderson, R.K. Time-Domain Fluorescence Lifetime Imaging Techniques Suitable for Solid-State Imaging Sensor Arrays. *Sensors* **2012**, *12*, 5650–5669, doi:10.3390/s120505650.
12. Antolovic, I.; Burri, S.; Hoebe, R.; Maruyama, Y.; Bruschini, C.; Charbon, E. Photon-Counting Arrays for Time-Resolved Imaging. *Sensors* **2016**, *16*, 1005, doi:10.3390/s16071005.
13. Shen, Q.; Liao, S.; Liu, S.; Wang, J.; Liu, W.; Peng, C.; An, Q. An FPGA-Based TDC for Free Space Quantum Key Distribution. *IEEE Trans. Nucl. Sci.* **2013**, *60*, 3570–3577, doi:10.1109/TNS.2013.2280169.
14. Homulle, H.; Visser, S.; Charbon, E. A Cryogenic 1 GSa/s, Soft-Core FPGA ADC for Quantum Computing Applications. *IEEE Trans. Circuits Syst. Regul. Pap.* **2016**, *63*, 1854–1865, doi:10.1109/TCSI.2016.2599927.
15. Vornicu, I.; Carmona-Galán, R.; Rodríguez-Vázquez, Á. Compensation of PVT Variations in ToF Imagers with In-Pixel TDC. *Sensors* **2017**, *17*, 1072, doi:10.3390/s17051072.
16. Ronchini Ximenes, A.; Padmanabhan, P.; Charbon, E. Mutually Coupled Time-to-Digital Converters (TDCs) for Direct Time-of-Flight (dTOF) Image Sensors ‡. *Sensors* **2018**, *18*, 3413, doi:10.3390/s18103413.
17. Akiba, K.; Ronning, P.; van Beuzekom, M.; van Beveren, V.; Borghi, S.; Boterenbrood, H.; Buytaert, J.; Collins, P.; Dosil Suárez, A.; Dumps, R.; et al. The Timepix Telescope for high performance particle tracking. *Nucl. Instrum. Methods Phys. Res. Sect. Accel. Spectrometers Detect. Assoc. Equip.* **2013**, *723*, 47–54, doi:10.1016/j.nima.2013.04.060.
18. Jeromel, L.; Siketić, Z.; Ogrinc Potočnik, N.; Vavpetič, P.; Rupnik, Z.; Bučar, K.; Pelicon, P. Development of mass spectrometry by high energy focused heavy ion beam: MeV SIMS with 8 MeV Cl7+ beam. *Nucl. Instrum. Methods Phys. Res. Sect. B Beam Interact. Mater. At.* **2014**, *332*, 22–27, doi:10.1016/j.nimb.2014.02.022.
19. Shin, S.; Jung, Y.; Kweon, S.-J.; Lee, E.; Park, J.-H.; Kim, J.; Yoo, H.-J.; Je, M. Design of Reconfigurable Time-to-Digital Converter Based on Cascaded Time Interpolators for Electrical Impedance Spectroscopy. *Sensors* **2020**, *20*, 1889, doi:10.3390/s20071889.

20. Kao, S.-K.; Hsieh, Y.-H.; Cheng, H.-C. An All-digital DLL with Duty-cycle Correction Using Reusable TDC. *Int J Circuit Theory Appl* **2016**, *44*, 1055–1070, doi:10.1002/cta.2124.
21. Ho, C.-R.; Chen, M.S.-W. 10.5 A digital PLL with feedforward multi-tone spur cancelation loop achieving <−73dBc fractional spur and <−110dBc Reference Spur in 65nm CMOS. In Proceedings of the 2016 IEEE International Solid-State Circuits Conference (ISSCC); 2016; pp. 190–191.
22. Staszewski, R.B.; Muhammad, K.; Leipold, D.; and; Wallberg, J.L.; Fernando, C.; Maggio, K.; Staszewski, R.; Jung, T.; John, and S.; et al. All-digital TX frequency synthesizer and discrete-time receiver for Bluetooth radio in 130-nm CMOS. *IEEE J. Solid-State Circuits* **2004**, *39*, 2278–2291, doi:10.1109/JSSC.2004.836345.
23. Cao, Y.; Cock, W.D.; Steyaert, M.; Leroux, P. Design and Assessment of a 6 ps-Resolution Time-to-Digital Converter With 5 MGy Gamma-Dose Tolerance for LIDAR Application. *IEEE Trans. Nucl. Sci.* **2012**, *59*, 1382–1389, doi:10.1109/TNS.2012.2193598.
24. Chan, S.; Warburton, R.E.; Gariepy, G.; Leach, J.; Faccio, D. Non-line-of-sight tracking of people at long range. *Opt. Express* **2017**, *25*, 10109–10117, doi:10.1364/OE.25.010109.
25. Tamborini, D.; Franceschini, M.A.; Stephens, K.A.; Wu, M.M.; Farzam, P.; Siegel, A.M.; Shatrovoy, O.; Blackwell, M.; Boas, D.A.; Carp, S.A. Portable System for Time-Domain Diffuse Correlation Spectroscopy. *IEEE Trans. Biomed. Eng.* **2019**, *66*, 3014–3025, doi:10.1109/TBME.2019.2899762.
26. Wang, D.; Wang, L.; Gao, P.; Shi, R.; Zhu, L.; Peng, Q.; Li, Z.; Zhao, J.; Chen, T.; Li, F.; et al. Simultaneous in vivo measurements of the total hemoglobin, oxygen saturation, and tissue blood flow via hybrid near-infrared diffuse optical techniques. *AIP Adv.* **2019**, *9*, 065306, doi:10.1063/1.5095699.
27. Alayed, M.; Deen, M.J. Time-Resolved Diffuse Optical Spectroscopy and Imaging Using Solid-State Detectors: Characteristics, Present Status, and Research Challenges. *Sensors* **2017**, *17*, 2115, doi:10.3390/s17092115.
28. Rahkonen, T.; Kostamovaara, J.; Saynajakangas, S. Time interval measurements using integrated tapped CMOS delay lines. In Proceedings of the Proceedings of the 32nd Midwest Symposium on Circuits and Systems,; 1989; pp. 201–205 vol.1.
29. Kalisz, J.; Szplet, R.; Pasierbinski, J.; Poniecki, A. Field-programmable-gate-array-based time-to-digital converter with 200-ps resolution. *IEEE Trans. Instrum. Meas.* **1997**, *46*, 51–55, doi:10.1109/19.552156.
30. Machado, R.; Cabral, J.; Alves, F.S. Recent Developments and Challenges in FPGA-Based Time-to-Digital Converters. *IEEE Trans. Instrum. Meas.* **2019**, *68*, 4205–4221, doi:10.1109/TIM.2019.2938436.
31. Wang, Y.; Liu, C. A Nonlinearity Minimization-Oriented Resource-Saving Time-to-Digital Converter Implemented in a 28 nm Xilinx FPGA. *IEEE Trans. Nucl. Sci.* **2015**, *62*, 2003–2009, doi:10.1109/TNS.2015.2475630.
32. Wang, Y.; Liu, C. A 4.2 ps Time-Interval RMS Resolution Time-to-Digital Converter Using a Bin Decimation Method in an UltraScale FPGA. *IEEE Trans. Nucl. Sci.* **2016**, *63*, 2632–2638, doi:10.1109/TNS.2016.2606627.
33. Chen, Y.-H. Time Resolution Improvement Using Dual Delay Lines for Field-Programmable-Gate-Array-Based Time-to-Digital Converters with Real-Time Calibration. *Appl. Sci.* **2019**, *9*, 20, doi:10.3390/app9010020.
34. Won, J.Y.; Lee, J.S. Time-to-Digital Converter Using a Tuned-Delay Line Evaluated in 28-, 40-, and 45-nm FPGAs. *IEEE Trans. Instrum. Meas.* **2016**, *65*, 1678–1689, doi:10.1109/TIM.2016.2534670.
35. Song, Z.; Wang, Y.; Kuang, J. A 256-channel, high throughput and precision time-to-digital converter with a decomposition encoding scheme in a Kintex-7 FPGA. *J. Instrum.* **2018**, *13*, P05012–P05012, doi:10.1088/1748-0221/13/05/P05012.
36. Wang, Y.; Liu, C. A 3.9 ps Time-Interval RMS Precision Time-to-Digital Converter Using a Dual-Sampling Method in an UltraScale FPGA. *IEEE Trans. Nucl. Sci.* **2016**, *63*, 2617–2621, doi:10.1109/TNS.2016.2596305.
37. Lai, J.-C.; Hsu, T.-Y. Cost-Effective Time-to-Digital Converter Using Time-Residue Feedback. *IEEE Trans. Ind. Electron.* **2017**, *64*, 4690–4700, doi:10.1109/TIE.2017.2669883.

38. Wang, Y.; Kuang, J.; Liu, C.; Cao, Q. A 3.9-ps RMS Precision Time-to-Digital Converter Using Ones-Counter Encoding Scheme in a Kintex-7 FPGA. *IEEE Trans. Nucl. Sci.* **2017**, *64*, 2713–2718, doi:10.1109/TNS.2017.2746626.
39. Shen, Q.; Liu, S.; Qi, B.; An, Q.; Liao, S.; Shang, P.; Peng, C.; Liu, W. A 1.7 ps Equivalent Bin Size and 4.2 ps RMS FPGA TDC Based on Multichain Measurements Averaging Method. *IEEE Trans. Nucl. Sci.* **2015**, *62*, 947–954, doi:10.1109/TNS.2015.2426214.
40. Chaberski, D. Time-to-digital-converter based on multiple-tapped-delay-line. *Measurement* **2016**, *89*, 87–96, doi:10.1016/j.measurement.2016.03.065.
41. Amiri, A.M.; Boukadoum, M.; Khouas, A. A Multihit Time-to-Digital Converter Architecture on FPGA. *IEEE Trans. Instrum. Meas.* **2009**, *58*, 530–540, doi:10.1109/TIM.2008.2005080.
42. Chen, P.; Hsiao, Y.; Chung, Y.; Tsai, W.X.; Lin, J. A 2.5-ps Bin Size and 6.7-ps Resolution FPGA Time-to-Digital Converter Based on Delay Wrapping and Averaging. *IEEE Trans. Very Large Scale Integr. VLSI Syst.* **2017**, *25*, 114–124, doi:10.1109/TVLSI.2016.2569626.
43. Szplet, R.; Kalisz, J.; Jachna, Z. A 45 ps time digitizer with a two-phase clock and dual-edge two-stage interpolation in a field programmable gate array device. *Meas. Sci. Technol.* **2009**, *20*, 025108, doi:10.1088/0957-0233/20/2/025108.
44. Won, J.Y.; Kwon, S.I.; Yoon, H.S.; Ko, G.B.; Son, J.; Lee, J.S. Dual-Phase Tapped-Delay-Line Time-to-Digital Converter With On-the-Fly Calibration Implemented in 40 nm FPGA. *IEEE Trans. Biomed. Circuits Syst.* **2016**, *10*, 231–242, doi:10.1109/TBCAS.2015.2389227.
45. Sui, T.; Zhao, Z.; Xie, S.; Xie, Y.; Zhao, Y.; Huang, Q.; Xu, J.; Peng, Q. A 2.3-ps RMS Resolution Time-to-Digital Converter Implemented in a Low-Cost Cyclone V FPGA. *IEEE Trans. Instrum. Meas.* **2018**, 1–14, doi:10.1109/TIM.2018.2880940.
46. Chen, K.; Liu, S.; An, Q. A high precision time-to-digital converter based on multi-phase clock implemented within Field-Programmable-Gate-Array. *Nucl. Sci. Tech.* **2010**, *21*, 123–128, doi:10.13538/j.1001-8042/nst.21.123-128.
47. Song, Z.; Zhao, Z.; Yu, H.; Yang, J.; Zhang, X.; Sui, T.; Xu, J.; Xie, S.; Huang, Q.; Peng, Q. An 8.8 ps RMS Resolution Time-To-Digital Converter Implemented in a 60 nm FPGA with Real-Time Temperature Correction. *Sensors* **2020**, *20*, 2172, doi:10.3390/s20082172.
48. Kwiatkowski, P. Employing FPGA DSP blocks for time-to-digital conversion. *Metrol. Meas. Syst.* **2019**, doi:10.24425/MMS.2019.130570.
49. Qin, X.; Zhu, M.-D.; Zhang, W.-Z.; Lin, Y.-H.; Rui, Y.; Rong, X.; Du, J. A high resolution time-to-digital-convertor based on a carry-chain and DSP48E1 adders in a 28-nm field-programmable-gate-array. *Rev. Sci. Instrum.* **2020**, *91*, 024708, doi:10.1063/1.5141391.
50. Kalisz, J. Review of methods for time interval measurements with picosecond resolution. *Metrologia* **2003**, *41*, 17–32, doi:10.1088/0026-1394/41/1/004.
51. Wu, J.; Shi, Z. The 10-ps wave union TDC: Improving FPGA TDC resolution beyond its cell delay. In Proceedings of the 2008 IEEE Nuclear Science Symposium Conference Record; 2008; pp. 3440–3446.
52. Bayer, E.; Traxler, M. A High-Resolution (<10 ps RMS) 48-Channel Time-to-Digital Converter (TDC) Implemented in a Field Programmable Gate Array (FPGA). *IEEE Trans. Nucl. Sci.* **2011**, *58*, 1547–1552, doi:10.1109/TNS.2011.2141684.
53. Pan, W.; Gong, G.; Li, J. A 20-ps Time-to-Digital Converter (TDC) Implemented in Field-Programmable Gate Array (FPGA) with Automatic Temperature Correction. *IEEE Trans. Nucl. Sci.* **2014**, *61*, 1468–1473, doi:10.1109/TNS.2014.2320325.
54. Wu, J. Several Key Issues on Implementing Delay Line Based TDCs Using FPGAs. *IEEE Trans. Nucl. Sci.* **2010**, *57*, 1543–1548, doi:10.1109/TNS.2010.2045901.

55. Chen, H.; Li, D.D. Multichannel, Low Nonlinearity Time-to-Digital Converters Based on 20 and 28 nm FPGAs. *IEEE Trans. Ind. Electron.* **2019**, *66*, 3265–3274, doi:10.1109/TIE.2018.2842787.
56. Dutton, N.; Vergote, J.; Gnecchi, S.; Grant, L.; Lee, D.; Pellegrini, S.; Rae, B.; Henderson, R. Multiple-event direct to histogram TDC in 65nm FPGA technology. In Proceedings of the 2014 10th Conference on Ph.D. Research in Microelectronics and Electronics (PRIME); 2014; pp. 1–5.
57. Knittel, G. A Novel Encoder for TDCs. In Proceedings of the Applied Reconfigurable Computing; Hochberger, C., Nelson, B., Koch, A., Woods, R., Diniz, P., Eds.; Springer International Publishing, 2019; pp. 48–57.
58. Liu, C.; Wang, Y. A 128-Channel, 710 M Samples/Second, and Less Than 10 ps RMS Resolution Time-to-Digital Converter Implemented in a Kintex-7 FPGA. *IEEE Trans. Nucl. Sci.* **2015**, *62*, 773–783, doi:10.1109/TNS.2015.2421319.
59. Xilinx UltraScale Architecture Configurable Logic Block User Guide (UG574) Available online: https://www.xilinx.com/support/documentation/user_guides/ug574-ultrascale-clb.pdf.
60. Wu, J. On-Chip processing for the wave union TDC implemented in FPGA. In Proceedings of the 2009 16th IEEE-NPSS Real Time Conference; 2009; pp. 279–282.
61. Wang, S.; Wu, J.; Yao, S.; Chang, W. A Field-Programmable Gate Array (FPGA) TDC for the Fermilab SeaQuest (E906) Experiment and Its Test with a Novel External Wave Union Launcher. *IEEE Trans. Nucl. Sci.* **2014**, *61*, 3592–3598, doi:10.1109/TNS.2014.2362883.
62. Lusardi, N.; Garzetti, F.; Geraci, A. The role of sub-interpolation for Delay-Line Time-to-Digital Converters in FPGA devices. *Nucl. Instrum. Methods Phys. Res. Sect. Accel. Spectrometers Detect. Assoc. Equip.* **2019**, *916*, 204–214, doi:10.1016/j.nima.2018.11.100.
63. Wang, Y.; Zhou, X.; Song, Z.; Kuang, J.; Cao, Q. A 3.0-ps rms Precision 277-MSamples/s Throughput Time-to-Digital Converter Using Multi-Edge Encoding Scheme in a Kintex-7 FPGA. *IEEE Trans. Nucl. Sci.* **2019**, *66*, 2275–2281, doi:10.1109/TNS.2019.2938571.
64. Szplet, R.; Kwiatkowski, P.; Jachna, Z.; Różyc, K. An Eight-Channel 4.5-ps Precision Timestamps-Based Time Interval Counter in FPGA Chip. *IEEE Trans. Instrum. Meas.* **2016**, *65*, 2088–2100, doi:10.1109/TIM.2016.2564038.
65. Wu, J. Uneven bin width digitization and a timing calibration method using cascaded PLL. In Proceedings of the 2014 19th IEEE-NPSS Real Time Conference; 2014; pp. 1–4.
66. Szymanowski, R.; Szplet, R.; Kwiatkowski, P. Quantization error in precision time counters. *Meas. Sci. Technol.* **2015**, *26*, 075002, doi:10.1088/0957-0233/26/7/075002.
67. Balla, A.; Mario Beretta, M.; Ciambrone, P.; Gatta, M.; Gonnella, F.; Iafolla, L.; Mascolo, M.; Messi, R.; Moricciani, D.; Riondino, D. The characterization and application of a low resource FPGA-based time to digital converter. *Nucl. Instrum. Methods Phys. Res. Sect. Accel. Spectrometers Detect. Assoc. Equip.* **2014**, *739*, 75–82, doi:10.1016/j.nima.2013.12.033.
68. Szplet, R.; Szymanowski, R.; Sondej, D. Measurement Uncertainty of Precise Interpolating Time Counters. *IEEE Trans. Instrum. Meas.* **2019**, *68*, 4348–4356, doi:10.1109/TIM.2018.2886940.
69. Zhao, L.; Hu, X.; Liu, S.; Wang, J.; Shen, Q.; Fan, H.; An, Q. The Design of a 16-Channel 15 ps TDC Implemented in a 65 nm FPGA. *IEEE Trans. Nucl. Sci.* **2013**, *60*, 3532–3536, doi:10.1109/TNS.2013.2280909.
70. Fishburn, M.; Menninga, L.H.; Favi, C.; Charbon, E. A 19.6 ps, FPGA-Based TDC With Multiple Channels for Open Source Applications. *IEEE Trans. Nucl. Sci.* **2013**, *60*, 2203–2208, doi:10.1109/TNS.2013.2241789.
71. Torres, J.; Aguilar, A.; García-Olcina, R.; Martı´nez, P.A.; Martos, J.; Soret, J.; Benlloch, J.M.; Conde, P.; González, A.J.; Sánchez, F. Time-to-Digital Converter Based on FPGA With Multiple Channel Capability. *IEEE Trans. Nucl. Sci.* **2014**, *61*, 107–114, doi:10.1109/TNS.2013.2283196.
72. Szplet, R.; Sondej, D.; Grzęda, G. High-Precision Time Digitizer Based on Multiedge Coding in Independent Coding Lines. *IEEE Trans. Instrum. Meas.* **2016**, *65*, 1884–1894, doi:10.1109/TIM.2016.2555218.

73. Chen, H.; Zhang, Y.; Li, D.D. A Low Nonlinearity, Missing-Code Free Time-to-Digital Converter Based on 28-nm FPGAs With Embedded Bin-Width Calibrations. *IEEE Trans. Instrum. Meas.* **2017**, *66*, 1912–1921, doi:10.1109/TIM.2017.2663498.
74. Qin, X.; Wang, L.; Liu, D.; Zhao, Y.; Rong, X.; Du, J. A 1.15-ps Bin Size and 3.5-ps Single-Shot Precision Time-to-Digital Converter With On-Board Offset Correction in an FPGA. *IEEE Trans. Nucl. Sci.* **2017**, *64*, 2951–2957, doi:10.1109/TNS.2017.2768082.
75. Lusardi, N.; Garzetti, F.; Geraci, A. Digital instrument with configurable hardware and firmware for multi-channel time measures. *Rev Sci Instrum* **2019**, 14.
76. UltraScale Architecture Libraries Guide (UG974) Available online: https://www.xilinx.com/support/documentation/sw_manuals/xilinx2014_1/ug974-vivado-ultrascale-libraries.pdf.
77. Designing with Low-Level Primitives User Guide Available online: https://www.intel.com/content/dam/www/programmable/us/en/pdfs/literature/ug/ug_low_level.pdf.
78. Kurtti, S.; Nissinen, J.; Kostamovaara, J. A Wide Dynamic Range CMOS Laser Radar Receiver With a Time-Domain Walk Error Compensation Scheme. *IEEE Trans. Circuits Syst. Regul. Pap.* **2017**, *64*, 550–561, doi:10.1109/TCSI.2016.2619762.
79. Zhang, M.; Wang, H.; Liu, Y. A 7.4 ps FPGA-Based TDC with a 1024-Unit Measurement Matrix. *Sensors* **2017**, *17*, 865, doi:10.3390/s17040865.